\documentclass[aps,pre,reprint,superscriptaddress]{revtex4-2}
\usepackage{amssymb}
\usepackage{amsmath,bm}
\usepackage{graphicx,color}
\usepackage[dvipsnames]{xcolor}
\usepackage{bbm}
\usepackage[bottom]{footmisc}
\usepackage{lipsum}
\usepackage{multirow}
\usepackage{soul}

\newcommand{\AM}[1]{\textcolor{black}{{#1}}}

\newcommand{\Rev}[1]{\textcolor{black}{{#1}}}
\usepackage[normalem]{ulem}
\usepackage{hyperref}
\makeatletter
\def\maketitle{
	\@author@finish
	\title@column\titleblock@produce
	\suppressfloats[t]}
\makeatother
\hypersetup{
    pdftoolbar=true,
    pdfmenubar=true,
    bookmarksopen=true,
    bookmarksnumbered=true,
    breaklinks=true,
    linktocpage,
    colorlinks=true,
    linkcolor={blue!95!black},
    citecolor={teal!70!black},
    urlcolor={blue!95!black},
}
 \newcommand{\beginsupplement}{%
	\setcounter{table}{0}
	\renewcommand{\thetable}{S\arabic{table}}%
	\setcounter{figure}{0}
	\renewcommand{\thefigure}{S\arabic{figure}}%
	\setcounter{equation}{0}
	\renewcommand{\theequation}{S\arabic{equation}}%
}

\begin{document}
\title{Finite Disorder Critical Point in Brittle-to-Ductile Transition in Amorphous Solids with Aspherical Impurities}

\author{Anoop Mutneja}
\email{anoop.mutneja2011@gmail.com} 

\affiliation{Department of Materials Science and Engineering, University of Illinois, Urbana, IL 61801, USA}
\affiliation{Materials Research Laboratory, University of Illinois, Urbana, IL, 61801, USA}

\author{Bhanu Prasad Bhowmik}
\email{bhowmikbhanuprasad592@gmail.com} 

\affiliation{
School of Engineering, The University of Edinburgh,
King’s Buildings, Edinburgh EH9 3FG, United Kingdom}
\author{Smarajit Karmakar}
\email{smarajit@tifrh.res.in} 
\affiliation{
Tata Institute of Fundamental Research, 36/P, Gopanpally Village, Serilingampally Mandal,Ranga Reddy District,
Hyderabad, Telangana 500107, India}

\begin{abstract}
Enhancing the mechanical properties of amorphous solids is crucial for material design, with microalloying being a common but not well-understood method. Using extensive molecular dynamics simulations, we investigate the effect of impurity particles on the yielding transition of amorphous solids in the context of brittle-ductile transition with microalloying. Spherical impurities larger than the constitutive particles enhance the system's mechanical stability, leading to a higher yield strain and increased brittleness. Much more potent effects are observed for rod-shaped impurities of the same size as the spherical impurities, with an aspect ratio slightly larger than one, which primarily introduce rotational degrees of freedom into the system. However, as the aspect ratio increases, their rotational degrees of freedom decrease, causing a more brittle yielding with more localized shear band formation. It is remarkable to see how freezing the rotational degrees of freedom can create extremely brittle, yet remarkably stable amorphous solids. Enhancing brittleness through higher concentrations of aspherical impurities presents an intriguing opportunity to explore the ductile-to-brittle transition, which is easily accessible in experiments, particularly in colloidal experiments. Our thorough finite-size scaling analysis has revealed a compelling suggestion of a finite-disorder critical point: a boundary between ductile and brittle behaviors.
\end{abstract}



\maketitle

\noindent{\bf \large Introduction: }The investigation of mechanical failure in amorphous solids holds paramount significance due to its extensive applications in both industrial and everyday life. While the mechanical behavior of crystalline solids subjected to external deformation is comprehensively elucidated in terms of defects associated with their structure, such defects cannot be precisely delineated in amorphous solids due to the absence of long-range structural order. As a result, despite rigorous research efforts~\cite{Barrat2011, Nicolas2018, Bonn2017, GhoshJOR2023}, the mechanical response of amorphous solids continues to elude comprehensive understanding.

Under external deformation, amorphous solids exhibit a combination of elastic and plastic responses~\cite{HKLPPRE2011, MaloneyLemaitrePRE2006, MaloneyLemaitrePRL2004, SmarajitLernerItamarJacquesPRE, GhoshJOR2023, mutneja2025}. As strain ($\gamma$) increases, stress ($\sigma_{xy}$) undergoes nonlinear growth, attributed to stress drops (plastic events) resulting from the irreversible rearrangement of constituent particles. In the initial stages of straining a freshly prepared sample, plastic events are small and spatially localized~\cite{HKLPPRE2011}, with a quadrupolar structure in the displacement field, well known in the literature from exact results in the Eshelby inclusion problem~\cite{Eshelby1957}. However, with increasing strain, their size and frequency amplify. Consequently, beyond a certain threshold of strain, stress ceases to increase, marking the onset of mechanical failure or fluidization of the solid. This phenomenon is recognized as the yielding transition of materials. Despite decades of research, a clear understanding of how an amorphous solid yields remains elusive. One set of studies suggests it to be a non-equilibrium phase transition~\cite{Nandi2016, Ozawa2018, WyartPRENonEqPhaseTran, Jaiswal2016, Parisi2017}, while other studies argue it is purely a dynamic crossover~\cite{Kobelev2005, SuSePNAS, GhoshJOR2023, mutneja2025}. Depending on the nature of the failure, one can distinguish between “ductile” and “brittle” yielding. In the former case, material flow is a gradual process facilitated by the proliferation of plasticity in the system, such as the flow of foam or various pastes~\cite{LaurisdenFoamPRL2002}. In the latter case, the system experiences catastrophic failure via the sudden formation of a system-spanning shear band instability, as observed in the breaking of metallic glass~\cite{GREER201371}. \Rev{However, it is not completely clear whether the same material can exhibit either form of yielding, depending on the sample’s stability~\cite{Ozawa2018, RossiTarju2022, Singh2020} or size~\cite{Fielding2020PRL, FieldingPRR, LernerBrittle, GreerScience}. }

\Rev{Recent studies suggest that the nature of the yielding transition is probably controlled by the inherent disorder strength of the samples~\cite{Ozawa2018, RossiTarju2022}. It was argued that brittle yielding is a nonequilibrium first-order transition (or spinodal) \cite{RainonePRL2015, UrbaniPRL2017,Jaiswal2016,Parisi2017,Ozawa2018} , exhibiting a discontinuous stress drop at macroscopic lengthscale, while ductile yielding shows a continuous stress-strain relationship reflecting gradual plastic softening. Under athermal quasistatic (AQS) conditions, these behaviors are separated by a critical point, suggesting a brittle-to-ductile transition akin to an athermally driven random-field Ising model (RFIM)\cite{Nandi2016,rossi2023far-from-equilibrium-5f0}. However, some research \cite{Fielding2020PRL, FieldingPRR, LernerBrittle} challenges this view, arguing that yielding is always brittle in large enough samples, regardless of their stability, indicating that the brittle-to-ductile transition may not even exist in the thermodynamic limit. Molecular simulations and large scale elastoplastic model simulation in which the transition is probed mostly by altering the stability of the samples via thermal annealing provided partial support, but conclusions remain uncertain.  While the size effect cannot be significantly overcome in molecular simulations, approaching this transition by controlling the inherent disorder strength through alternative methods—accessible in a broader range of experimental systems, including athermal systems like colloidal solids where thermal annealing is not applicable—might provide more conclusive evidence for the existence of this non-equilibrium transition. This is suggested in a recent study \cite{mutnejaBPB2025}, where disorder is adjusted through particle pinning. While particle pinning can be achieved in colloidal glass experiments\cite{gokhale2014growing}, it remains a challenge for molecular glasses. However, the potential for achieving this through soft pinning has been explored \cite{das2023soft, anwar2024exploring, meenakshi2024effect}. Therefore, identifying new protocols to investigate this transition—particularly one that is accessible in colloidal experiments—could have significant implications for the field. It is needless to reiterate that a better understanding of the brittle-ductile transition is fundamentally important for both scientific advancement and industrial applications.} 

\Rev{In a different context, the preparation of a highly stable amorphous solid along with the optimization of its mechanical strength is one of the most important areas of ongoing research.} From the perspective of numerical studies, amorphous solids are typically prepared by fast cooling of a model glass former, where a lower cooling rate produces a more mechanically stable sample~\cite{ItamarEranJoyPRE2013}. However, even the lowest accessible cooling rate results in poorly annealed states, making it difficult to study realistic systems. Highly stable glassy configurations, known as ultrastable glasses, can be prepared using the swap Monte Carlo method in computer simulations~\cite{MCGrigeraParisi, SmarajitSwapTernary} in polydisperse systems~\cite{Ozawa2018}, and through cyclic deformation~\cite{Bhowmik2020, Bhaumik2021}. While the swap Monte Carlo method is not readily applicable to experiments, cyclic shear protocols may not guarantee kinetic stability or consistently produce well-annealed glasses~\cite{PhysRevApplied.19.024004, Bhaumik2021, LudovicCyclicShear}. Recently, ultrastable glasses have been obtained using random bonding~\cite{Ozawa2023}, which is considered a promising technique for generating realistic stable glasses in patchy colloids, in addition to vapor deposition methods~\cite{EdigerVD}.

\Rev{In parallel to the advancements in producing stable amorphous solids, an equally important aspect of research is gaining control over the load bearing capacity of these materials.} For decades, engineers have explored ways to improve the ability of amorphous solids to sustain external loading. One widely used method is microalloying, which involves adding small amounts of other elements (impurities) to the pure sample~\cite{Garrett2012, Gonzlez2015}. While there are many experimental examples showing that microalloying~\cite{PhysRevLett.99.135502, Garrett2012} can enhance the yield strain, the microscopic mechanisms, behind this improvement remain elusive.  Recent studies have also focused on the role of impurities in the mechanical response of amorphous solids, particularly through particle pinning~\cite{Bhowmik2019, Corrado2019, mutnejaBPB2025}. It has been shown that the presence of pinned particles prevents plastic events and delays the yielding transition. Moreover, random pinning leads to a transition from brittle to ductile yielding as the pinned particle concentration varies~\cite{mutnejaBPB2025}. \Rev{Building upon these findings, we numerically explore the implications of incorporating bulkier spherical and aspherical impurities (soft pinning centers) into amorphous solids on their yielding transition, thereby seeking to further elucidate the role of structure disorder in their failure processes.}

\begin{figure}
\includegraphics[width=.470\textwidth]{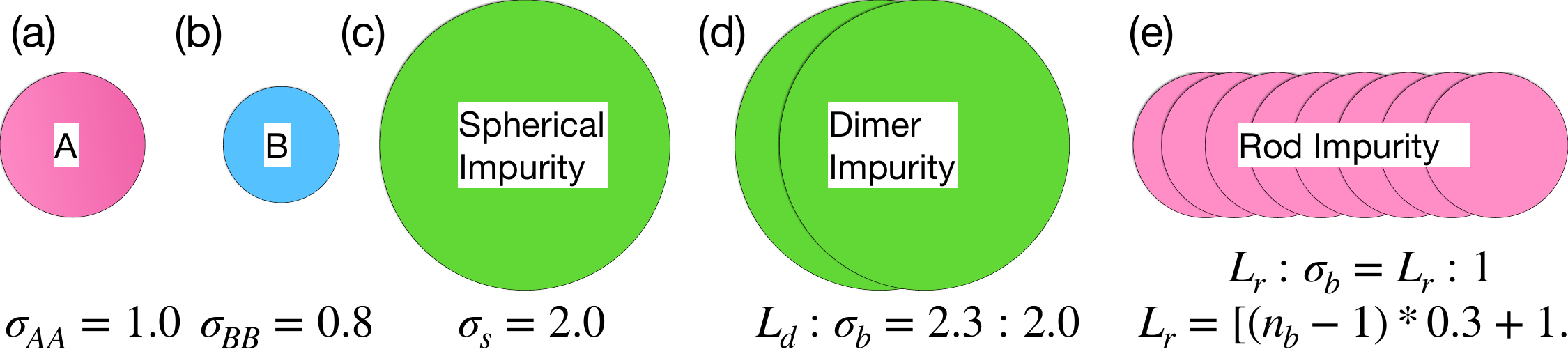}
\caption{\textbf{Components of the system: }(a, b) Particle type ‘A’ and ‘B’ of the parent KA system. (c) The spherical impurity with twice the diameter of the particle ‘A’. It has a variable number fraction of $c_s$ in the system. (d) A dimer made with larger particles and having a slight asphericity; we have varied their number fraction ($c_d$) in the system. (e) Rods with larger aspect ratio, made by attaching A-type particles. We studied the systems with different aspect ratio rods, with a fixed number fraction of $c_r=0.1$.}
\label{fig1}
\end{figure}

Using computer simulations of a binary glass-forming system, we investigate the effect of impurities on the yielding transition of amorphous solids under AQS straining conditions \Rev{with an aim to decipher the existence of a non-equilibrium Brittle-Ductile phase transition}. We consider both spherical (Fig.~\ref{fig1}(c)) and aspherical (Fig.~\ref{fig1}(d, e)) impurities. For spherical impurities, we add a third type of particle into the glass matrix, which has a larger diameter than the constituent particles. For aspherical impurities, we introduce rod-shaped particles. The advantage of using rod-shaped impurities over spherical ones is that they introduce rotational degrees of freedom (rDoF) into the system, providing additional pathways for stress relaxation under mechanical loading. We study the changes in the mechanical response of amorphous solids due to the presence of impurities by varying the concentration for spherical impurities, and both the concentration and the aspect ratio for aspherical impurities, and connect it with the structural stability of the samples.

As a preview, we find that the inclusion of large spherical impurities leads to reduced plasticity at smaller strains, a higher yield strain, and more brittle-type failure. We connect these effects with enhanced structural order, quantified by a structural order parameter defined in Ref.~\cite{Tong2019}. In the case of aspherical impurities, rod-shaped particles with aspect ratios close to one (i.e., small asphericity) exhibit a similar but more pronounced response. The system can sustain a significantly larger load compared to systems with spherical impurities. We establish that this enhancement is linked to the additional rotational degrees of freedom (rDoF) present in  the system of rod-shaped particles, but absent in the spherical case (Fig.~\ref{fig1}(b)). We further investigate the effect of increasing the aspect ratio of rod-shaped particles by increasing the length of rods while keeping their radius fixed. We find that these longer rods, which have greater rotational inertia, lead to lower rotational diffusion, reduce the yield strain enhancement achieved, and induce a more brittle yielding response. Our study thus suggests that the presence of rDoF enhances the strain-bearing capacity of the system, whereas freezing the rDoF leads to an ultrastable-like mechanical response, consistent with recent findings~\cite{Ozawa2023}.To verify this, we examine the mechanical response of a system in which rod-shaped impurities are artificially kept rotationally frozen during the stress release process. This analysis confirms that the presence of rotationally immobile rod-shaped impurities results in ultrastable amorphous solids. \Rev{Finally, we used this new way to induce brittleness as a tool to study the ductile-brittle transition by increasing the concentration of rotationally frozen rods via detailed finite size scaling analysis. We further show that the transition occurs at a critical point corresponding to a non zero rod concentration in the thermodynamic limit.}

\Rev{To underscore the novelty and applicability of our findings, we briefly examine the significant potential our results present for soft glassy systems, particularly colloidal glasses. Recent advancements in a variety of imaging techniques have highlighted the critical importance of colloidal glass experiments \cite{NagaManasaNatPhys, gokhale2014growing, ChikkadiPRL2011, ChikkadiPRL2014} in elucidating the complex characteristics associated with glass transition, alongside the role of plasticity in the mechanical properties of amorphous solids. The identification of Eshelby-like plastic responses in colloidal glasses subjected to simple shear \cite{ChikkadiPRL2014} has created exciting opportunities for investigating yielding behaviors in amorphous solids, utilizing colloidal glasses as representative experimental models. Furthermore, numerous insightful conclusions regarding the dynamical and mechanical properties of amorphous solids, derived from computer simulations, have been corroborated by experimental observations in colloidal glass studies \cite{NagaManasaNatPhys, gokhale2014growing, ChikkadiPRL2011, ChikkadiPRL2014}. This validation indicates that colloidal glass experiments serve as an exemplary platform for testing various theoretical predictions within real-world contexts. Consequently, the exploration of the brittle-ductile transition in colloidal glasses warrants particular attention and could yield fascinating insights.}

\Rev{The primary obstacle in examining the brittle-ductile transition within colloidal systems lies in the challenge of accurately tuning the disorder level using current annealing methods. While molecular glasses can utilize temperature annealing or vapor deposition techniques to create ultra-stable glasses, these strategies prove unsuitable for colloidal samples. Although oscillatory shear can be applied to anneal colloidal solids, its limitations restrict significant tuning. Current research suggests that under optimal experimental conditions, colloidal glasses display ductile-like yielding, and brittle-like failures have yet to be observed in these materials. Our innovative strategy to provoke a brittle response in ductile materials through microalloying with elongated impurities may unlock the ability to induce brittleness even in soft solids. This method could represent one of the few viable pathways to exploring the brittle-ductile transition in colloidal glasses, especially as prior experiments with ellipsoidal colloidal particles have already demonstrated its feasibility \cite{mishraPNASellipsoids2014, mishraPRL2015ellipsoids}.}


\vskip +0.1in
\noindent{\bf \large System Details: } We simulate a system of Lennard-Jones (L-J) particles of two different types, with a number ratio of 80:20 in three dimensions (3D), known as the Kob-Andersen (KA) model glass former (3DKA)~\cite{KA}, and a number ratio of 65:35 in two dimensions (2D), known as the modified Kob-Andersen model glass former (2DmKA)~\cite{2dmKA}. The quiescent states are prepared by cooling the system from a high-temperature melt or by equilibrating the system at a given temperature, followed by quenching it to an inherent state in both cases. The inherent states are then subjected to external deformation using athermal quasistatic straining (AQS). Fig.\ref{fig1} shows the system constituents, including the binary system particles and the embedded impurities. The rod-shaped particles are made by attaching $n_b$ spherical particles at a fixed distance of 0.3 units; see Fig.\ref{fig1} (d) and (e). The structural order of the quiescent states with various impurities is characterized by a structural order parameter $\Theta$, developed in Ref.\cite{Tong2019}, based on the separation distance of the first nearest neighbors of a particle obtained by Voronoi tessellation \cite{PhysRevE.74.021306}. Details of the simulation, models and $\Theta$ are provided in the Models and Methods section.

\section*{Results}
\begin{figure}
\includegraphics[width=0.47\textwidth]{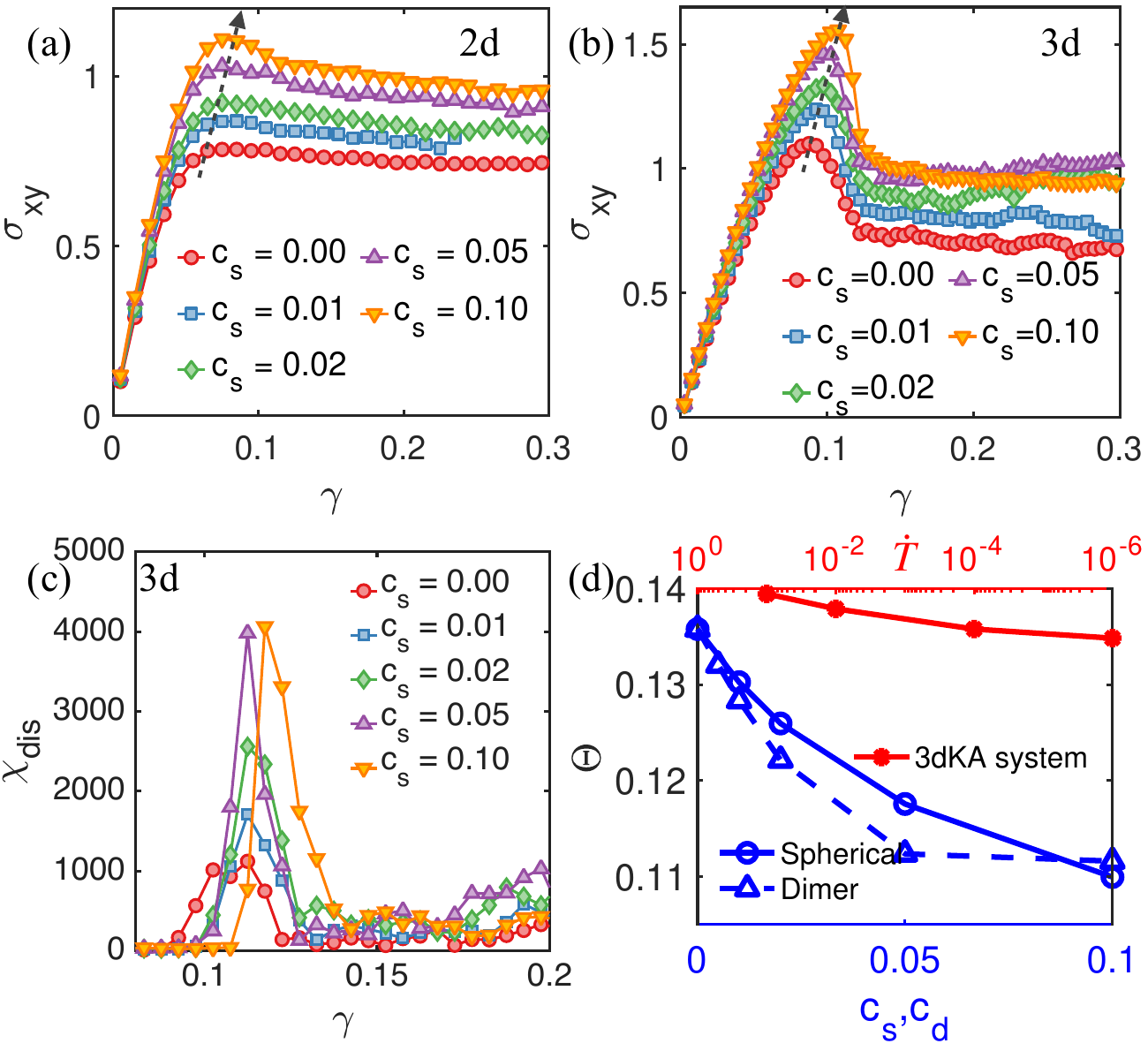}
\caption{{\textbf{Mechanical properties with spherical inclusions: }(a, b) Stress-strain curves for the 2dmKA and 3dKA systems with different number fractions of spherical impurities ($c_s$); the systematic shift in the yield point to higher strain values with increasing the concentration of ternary particles can be seen. Panel (c) shows the susceptibility plot (Eq.(\ref{eqn:eqn1})) for the 3d system with changing $c_s$. The shifted peaks and increased sharpness conclude the increase in yield point and emerging brittle-like behavior. (d) The structural order parameter is shown for the quiescent ($\gamma=0$) state of the system with different concentrations of inclusions (blue), along with its comparison with the states generated using different cooling rates (red). It shows the increased structural stability of the glass matrix with increasing $c_s$ and not so much with decreasing cooling rate.}}
\label{fig2}
\end{figure}

\noindent{\bf \large Tuning mechanical strength with spherical impurities:}
First, we investigate the bulk mechanical behavior of the system by incorporating spherical impurities (Fig~\ref{fig1}(c)). Generally, such impurities have two adjustable parameters: their size and the strength of their interaction with the system’s constituent particles. While the latter has been studied in the past~\cite{Gendelman2014}, we focus on the effect of size by introducing larger impurities with a diameter of $\sigma_s=2\sigma_{AA}$, for a range of number fractions $c_s\in0-0.1$, into the system of total particles (constitutive particles plus impurities) $N_T=100000$. We conduct a strain-controlled deformation on the 2DmKA and 3DKA glass-forming systems, with the stress acting as the observable. Fig.~\ref{fig2} (a) and (b) illustrate the stress-strain curves for both models, respectively.

The results clearly indicate an increase in the yield strain ($\gamma_Y$) and the shear modulus ($\mu$) with increasing $c_s$. The enhancement of mechanical strength is also accompanied by a prominent stress overshoot, especially for the 3DKA model, along with an increase in steady-state stress values. Moreover, the yielding process seems to become more brittle type. Generally, the nature of the yielding transition can be characterized by computing the disconnected susceptibility,
\begin{equation}
\chi_{dis}(\gamma)=N_T \left(\langle\sigma_{xy}^2\rangle-\langle\sigma_{xy}\rangle^2\right).
\label{eqn:eqn1}
\end{equation}
Here, $\langle ... \rangle$ represents ensemble averaging. Typically, $\chi_{dis}$ attains its maximum value, $\chi^p$, at $\gamma_Y$ due to the maximum occurrence of plastic events. A larger value of $\chi_{dis}^p$ and a narrower $\chi_{dis}$ -- $\gamma$ curve indicate a more brittle type failure~\cite{Ozawa2018}. In Fig.~\ref{fig2}(c), we show how the $\chi_{dis}$ -- $\gamma$ curves for 3D system to become more narrower and how $\chi_{dis}^p$ increases with increasing $c_s$, signifying the emergence of a more brittle character of failure in our system. The position of the peak also shifts to larger $\gamma$, further supporting the shift in yield strain due to the presence of impurities. Similar $\chi_{dis}-\gamma$ plots for 2D systems are provided in \textit{SI} (\textit{supporting information}) Fig.S1.

The mechanical response of systems with impurities indicates enhanced structural stability of the samples. The most common method to characterize the stability of an amorphous solid is the energy per particle, averaged over all the particles in the system. However, in our system, due to the addition of foreign particles, the energy scale across different concentrations might not remain comparable. Thus, to delve deeper into this observation, we rely on the structural order parameter~\cite{Tong2019}, denoted by $\Theta$ (Eq.[\ref{eqn:eqn4}]), which depends purely on the local structure rather than the interaction. A lower value of the structural order parameter indicates better stability of the system. We compute $\Theta$ for the quiescent (unstrained) state at all concentrations, as shown in Fig.~\ref{fig2}(d). We find that $\Theta$ decreases with increasing $c_s$, indicating enhanced structural stability and order, aligning with the observed more brittle behavior. For comparison, we include $\Theta$ for systems prepared using different cooling rates (red points), revealing their minimal impact on structural stability compared to the effect of large spherical impurities. We choose an impurity diameter of $2.0\sigma_{AA}$ because, on one hand, smaller impurities have a very mild effect, and on the other hand, systems with larger impurities cannot be equilibrated within the available simulation time scale due to their slow dynamics. 
\vskip +0.1in
\noindent{\bf \large Tuning mechanical strength using aspherical impurities:}
Although the results with spherical impurities are encouraging, we must admit that the improvement is minimal ($\gamma_Y\left(c_s=0\right)=0.09$ to $\gamma_Y\left(c_s=0.1\right)=0.107$ for the 3DKA system). Interestingly, studies in \cite{Bhowmik2019,Ozawa2023,LIU2023122052,mutnejaBPB2025} highlight the significant impact of eliminating degrees of freedom on the mechanical properties of amorphous solids. In contrast, we focus on altering the mechanical properties of amorphous solids brought about by introducing rotational degrees of freedom (rDoF) through rod-shaped aspherical impurities, as discussed in the following sections.
\vskip +0.1in
\noindent{\bf Effect of impurity concentration:}
\begin{figure*}
\includegraphics[width=15.65cm]{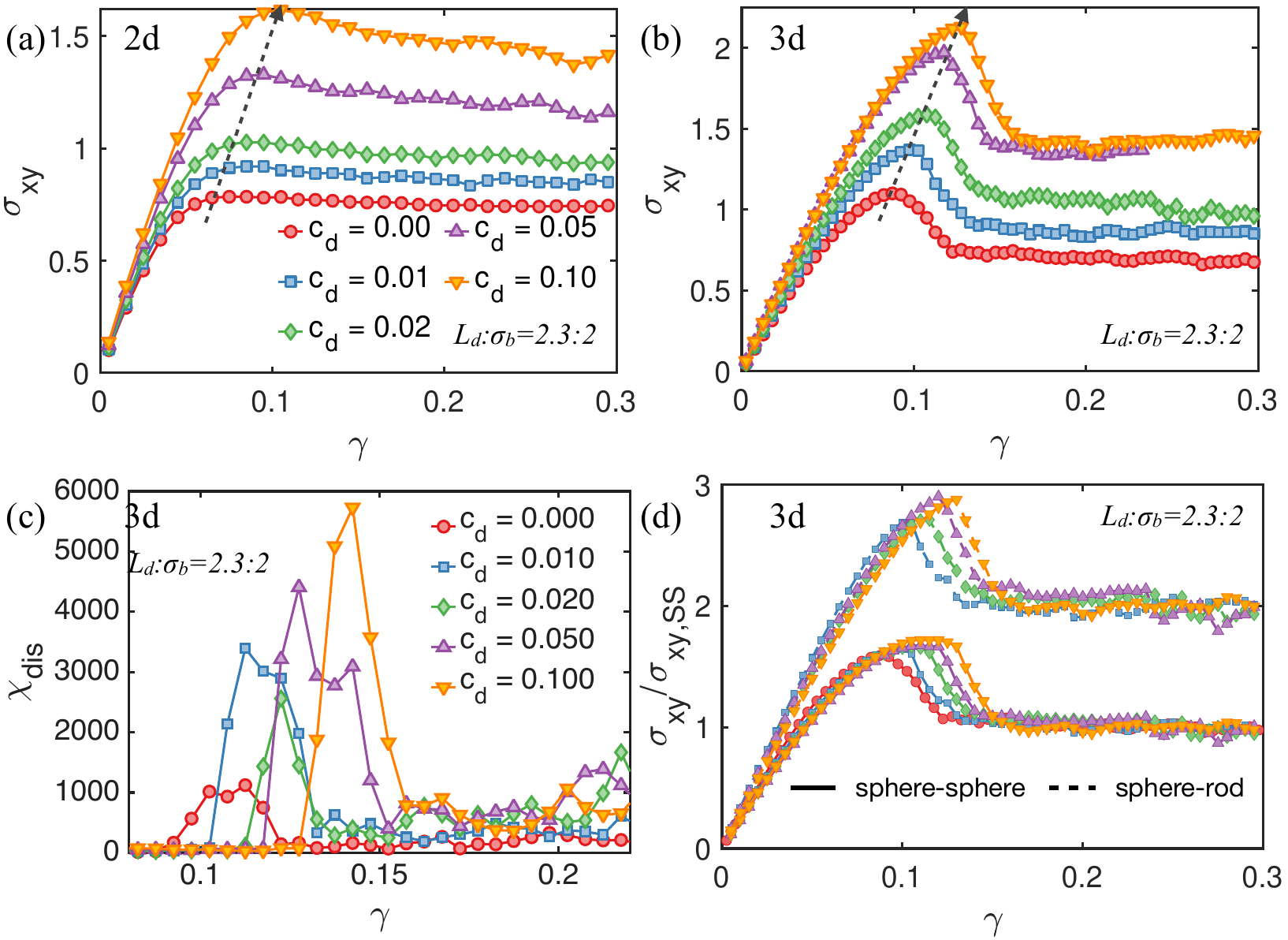}
\caption{{\textbf{Mechanical properties with aspherical impurities: Effect of impurity concentration} (a, b) Stress-strain curves for 2dmKA and 3dKA systems with different dimer concentrations $c_d$; the systematic shift in the yield point to higher strain values, almost 100\% improvement compare to the ternary system is seen; Panel (c) is the susceptibility plots for the three dimension system. The increase in peak height and the shift in peak to larger strain value with increasing $c_d$ supports the conclusion of increased yield strain. Panel (d) shows the stress contributions from sphere-sphere and sphere-rod bonds, normalized by the steady-state values. The rods-sphere pairs are making the system hold stress to larger strain values, where the sphere-sphere stress has already started to fall.
}}
\label{fig3}
\end{figure*}
To begin with, we examine the response of less aspherical rod-shaped impurities (Fig.\ref{fig1}(d)) for various impurity concentrations, $c_d$. The volume of these impurities is chosen to be similar to that of spherical impurities, aiming to impart a comparable influence on the system as the larger spherical impurities, as demonstrated by the structural order parameter in Fig.\ref{fig2}(d). However, compared to spherical impurities, they possess additional rotational degrees of freedom. These impurities are created by attaching two spheres (beads) with a diameter of $\sigma_b = 2.0\sigma_{AA}$, with a $77.7\%$ overlap, resulting in a length of $L_d = 2.3\sigma_{AA}$ and an aspect ratio of $L_d:\sigma_b = 2.3:2$. Given that these impurities consist solely of two beads, we designate them as dimers to differentiate them from another category of impurities comprising more than two beads with the same amount of overlap, described later in the paper. Various parameters of the dimer impurity system are denoted by the subscript ‘d’ unless stated otherwise. Fig.\ref{fig3}(a, b) shows the stress–strain curves for the studied 2D and 3D model systems. Notably, we observe a significant increase in yield strain for both 2D and 3D systems, accompanied by a substantial increase in the shear modulus. For instance, in the 3DKA model, $\gamma_Y$ increases by around $40\%$, from $\gamma_Y = 0.09$ for the pure system to $\gamma_Y = 0.127$ for $c_d = 0.1$, compared to around $18\%$ for the system with spherical impurities. Additionally, the stress overshoots become more pronounced with higher dimer concentrations. This systematic increase in both yield stress and strain with larger $c_d$ suggests that small dimer impurities render the system more stable and capable of withstanding greater loads than the pure system. This observation is further supported by the $\chi_{dis}$–$\gamma$ plots shown in Fig.\ref{fig3}(c) for 3D system (Fig.S2 in \textit{SI} for 2D system). While the increased value of $\chi_{dis}^p$ suggests larger stress drops, the spread of the $\chi_{dis}$–$\gamma$ curves remains more or less the same, indicating no significant change in the nature of the yielding.  We note that the thinner dimer impurities with aspect ratio $L_r:\sigma_b = 1.3:1$ (Fig.~\ref{fig4}(a)) give qualitatively similar results, although the changes in various mechanical properties are less prominent. We do not study dimers with a larger diameter, as equilibrating such systems poses a challenge.

To microscopically understand the role of rod impurities in enhancing the system’s stability and delaying the yielding transition, we analyzed the stress contributions from rod–sphere and sphere–sphere interactions at a microscopic level, as shown in Fig.\ref{fig3}(d). The data for all studied concentrations reveal that the stress contribution from sphere–sphere interactions begins to saturate at smaller strains than the system’s macroscopic yield strain, whereas the stress contribution from rod–sphere interactions continues to increase until the yield strain and then decreases sharply. This indicates that regions containing rods are more structurally stable and can sustain loads at higher strain values than those without rods. This also suggests that the onset of yielding is mainly controlled by rod–sphere interactions, providing a clear picture of delayed yielding with increasing $c_d$. A closer look at the single-ensemble stress–strain response (shown in \textit{SI} Fig.S3) reveals a similar pattern: rod–sphere pairs continue to store stress, while the stress in sphere–sphere pairs has already saturated, despite a one-to-one correspondence between their plastic events.

Fig.~\ref{fig2}(d) illustrates the enhanced structural stability of the parent glass matrix in terms of $\Theta$ as the dimer concentration increases, along with its variation with the concentration of spherical impurities and with different cooling rates for comparison. It is important to note that the structural stability introduced by both spherical and aspherical particles is similar in magnitude. This can be attributed to the fact that the volume of a rod is comparable to that of a spherical impurity. However, the substantial shift in the yield point for the aspherical case indicates a significant influence of rDoF on the yielding transition. By providing additional local pathways for dissipating internal stresses, these degrees of freedom allow the system to release extra stress sustain higher loads and store more stress, before a non-local shear banding event leads to failure. In the subsequent section, we explore systems containing rod-shaped impurities of varying lengths to test this assertion further, as such impurities with longer lengths inherently exhibit reduced rotational freedom.
\vskip +0.1in
\noindent{\bf  Effect of the length of rod impurity:}
\begin{figure*}[!htp]
\includegraphics[width=.980\textwidth]{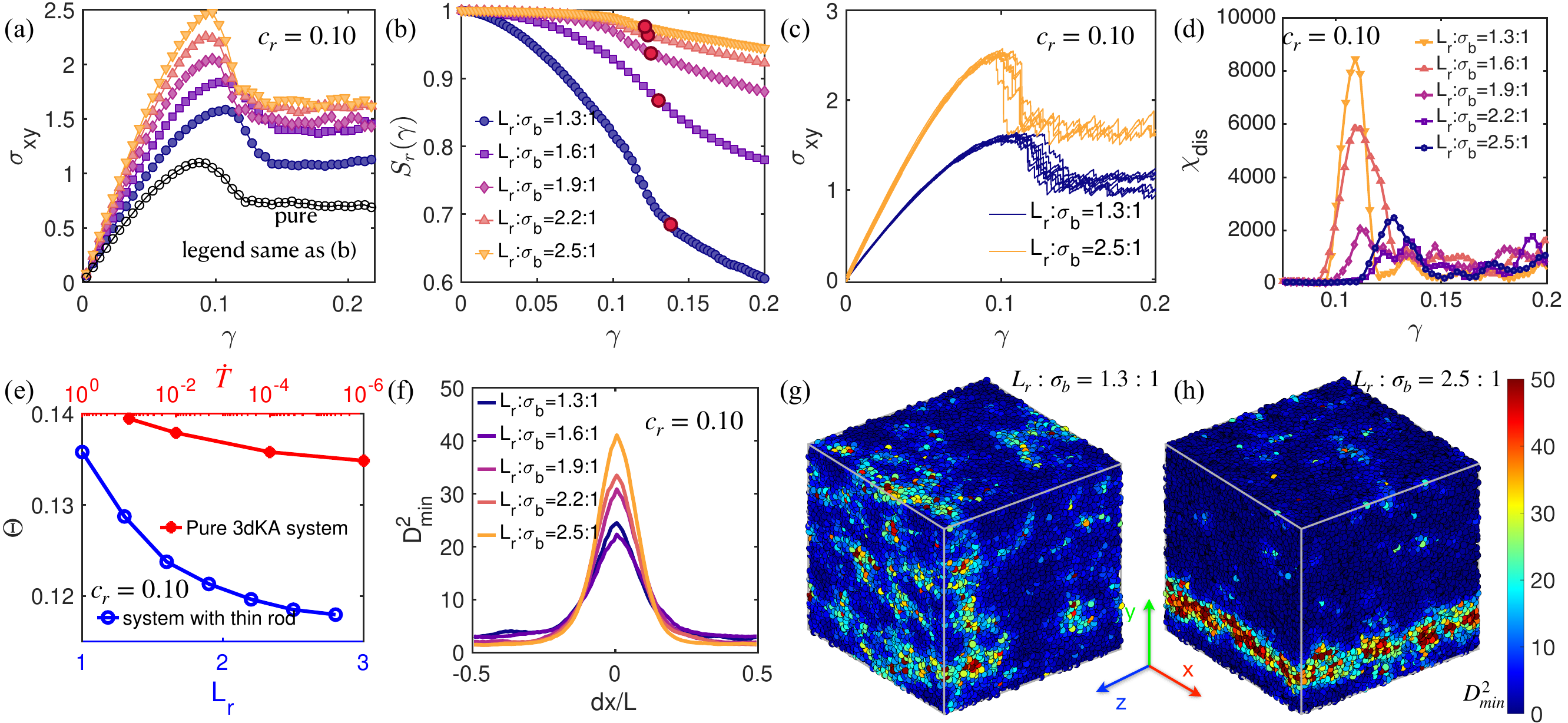}
\caption{\textbf{Mechanical properties with aspherical impurities: Effect of the length of rod impurity:} (a) The stress-strain curves for 3dKA system with different lengths of rods (Fig.~\ref{fig1}(e)). Firstly, the yield point shifts back with increasing rod length because of losing rDoF, which is shown in panel (b). Panel (b) shows the rotational de-correlation function for rods of different lengths with mechanical loading; the red points indicate the equal net $D_{min}^2$ \cite{Falk1998}. Secondly, Panel (a) suggests the increased brittle behavior with increasing rod length. It is further advocated in (c-h). Panel (c) shows brittle behavior in sample-to-sample stress-strain plots for systems with rods of length $L_r=2.5$ ($L_r=1.3$). The sharpness and increased peak height of $\chi_{dis}$ in (d) also conveys the emerging mechanical ultra-stability of the sample. (e) The structural order parameter decreases with increasing rod length in the system, implying increased structural stability. Panel (f) shows the non-affine displacement $D_{min}^2$ \cite{Falk1998}, averaged in strips perpendicular to the shear band (at $dx=0$). All these curves have the same area under the curve to ensure equal plasticity. The large displacement away from the shear band in systems with smaller rods and large peak value for longer rods again advocate the emerging brittle behavior. It is also clear from the $D_{min}^2$ maps (g, h) obtained at equal net displacements for systems with rods of lengths $L_r=1.3$ and $L_r=2.5$, respectively.}
\label{fig4}	
\end{figure*}
We now add longer rod-shaped impurities (more aspherical compared to dimer) in the amorphous matrix (see Fig.~\ref{fig1}(e)). These impurities have two or more than two beads and, for clarity, are identified as rods. The subscript ‘r’, if not mentioned otherwise, indicates various parameters associated with the rod impurity system. The diameter of each bead in the rod is kept to be same as the larger particle of the parent system,  $\sigma_r=\sigma_{AA}$. The reason to reduce the bead diameter from $2\sigma_{AA}$ to $\sigma_{AA}$ is that long rods with large diameters exhibit very slow dynamics and can not be equilibrated within the available simulation time scale. Even with a smaller bead, the maximum rod length we can simulate is $L_r=2.5\sigma_{AA}$. Fig.~\ref{fig4}(a) shows the averaged stress-strain curve for a 3dKA system with rod concentration $c_r=0.1$ of different $L_r$. One can clearly see that for samples with the rod length, $L_r=1.3$ has large $\gamma_Y$ compared to the pure samples (shown in black) due to the presence of rotational degrees of freedom, but the yield point is shifting back to lower values with increasing $L_r$, owing to the fact of reduced rotational relaxation of the longer rods. 

The degree of rotational relaxation can be characterized by the rotational relaxation function defined as: 
\begin{equation}
    S_r = \left\langle\frac{1}{N_r}\sum_{i=1}^{N_r}\hat{s}_i(\gamma).\hat{s}_i(0)\right\rangle
\end{equation}
and is shown in Fig.~\ref{fig4}(b). Here ${\hat{s}}_i$ is the orientation vector of the rod. A faster decay of $S_r$ suggests a better independent mobility of the rod’s orientation. We see $S_r$ decreases with $\gamma$ for all $L_r$ due to the non-affine motion originating from the plastic events. However, since systems with different $L_r$ do not suffer from the same amount of plasticity at the same $\gamma$, we choose $S_r$  at different $\gamma$ shown by the red points on the data where all the systems have the same amount of net non-affine displacement (in terms of total $D_{min}^2$ \cite{Falk1998} ). The higher values of $S_r$ at the point of the same plasticity with increasing $L_r$ advocate the reduced rotational motion of larger rods.

Another clear facet emerging from the average stress-strain curves is the increased brittle behavior with increasing rod length. It is also worth pointing out that the system with longer rods is dynamically slower. Therefore, the same preparation protocol would produce less annealed states for longer rod systems, which should smoothen the stress-strain curve. This implies that the observed brittleness is even stronger than what we obtain in this study because of the simulation difficulty. In Fig.~\ref{fig4}(c), the individual stress-strain curves of two different $L_r$ are compared. For $L_r=1.3$, the stress-strain curves exhibit several stress drops near the yielding transition. On the other hand, for $L_r=2.5$, the individual stress-strain curves are more abrupt, with larger stress drops close to $\gamma_Y$, indicating the emergence of shear band instability in the system. Fig.~\ref{fig4}(d) displays the large values of the $\chi_{dis}\left(\gamma\right)$ peak, which becomes narrower with increasing $L_r$, indicating an emerging mechanically brittle phase for longer rods. In panel (e), the variation of $\Theta$ in the quiescent state is presented as a function of rod length for samples prepared using the same cooling rate. $\Theta$ decreases with increasing $L_r$, suggesting better structural stability of the samples. 
\begin{figure*}
\centering
\includegraphics[width=0.99\textwidth]{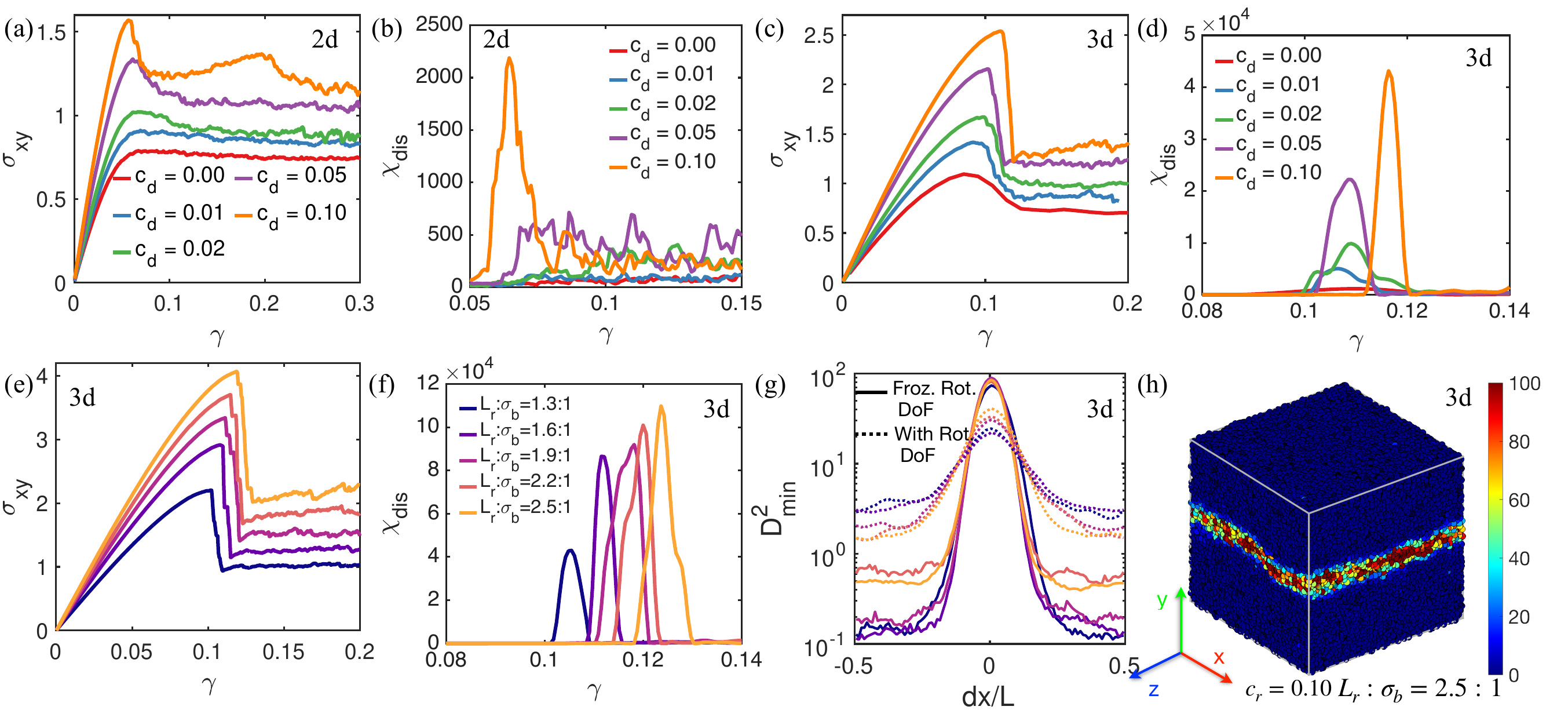}
\caption{{\textbf{Ultra-stability with frozen rDoF: Mechanical aspect:} (a, c) The averaged stress-strain curves for 2dmKA and 3dKA systems with rotationally frozen dimer (Fig.~\ref{fig1}(d)) inclusions. The yield strain does not change much, and the system becomes extremely brittle, as seen from the respective susceptibility plots in panels (b, d). Note for the 3dKA model, the stress-strain curves become nearly discontinuous at the yielding transition for the larger concentration of dimers. The effect is even enhanced for the system with more aspherical rods (panel (e, f)). Inset of (f) contains the averaged maximum stress drop ($\Delta\sigma_{max}$) that happened in the whole strain window. It is an order parameter for brittle behaviour \cite{Ozawa2018}, and its systematic increased magnitude with increasing rod concentration and rod length proves that the frozen rDoF leads to a highly brittle breakdown. 
Panel (g, h) shows the $D^2_{min}$ plots indicating the sharpness of the obtained shear band (see text for detailed discussion).}}
\label{fig5}
\end{figure*}

As a direct indicator of increased brittle behavior, Fig.~\ref{fig4}(f) presents the non-affine displacement, $D_{min}^2$ \cite{Falk1998}, averaged across slices perpendicular to the shear band and plotted against the distance from the center of the shear band for systems with rods of varying lengths. All the curves have similar areas under the curve to ensure the same amount of plasticity. The plots show that the non-affine displacement magnitude away from the shear band decreases with increasing rod length, indicating less spatially scattered plastic events. Furthermore, the peaked displacement for larger rods shows the formation of shear band. The same can be seen from $D_{min}^2$ map of the system with $L_r=1.3$ in Fig.~\ref{fig4}(g), and with $L_r=2.5$ in Fig.~\ref{fig4}(h). For a direct comparison, these two maps have an equal net $D_{min}^2$, and it is evident that the system with shorter rods has multiple plastic events that are spatially spread out, while the longer rods have a localized shear band. Thus, the emerging brittle behavior originates from the lack of rotational freedom, as we will see in the next section.
\vskip +0.1in
\section*{Ultra-stability with Frozen \lowercase{r}D\lowercase{o}F}
\vskip +0.1in
\noindent{\bf Mechanical aspect:} 
Our results presented in Fig.~\ref{fig4} demonstrate that with increasing rod length, the rotational diffusivity decreases significantly, indicating the possibility of complete arrest of rotation after a certain rod length. However, the effect of such long rods cannot be observed due to simulation limitations. Nonetheless, in a realistic system, it is possible to significantly reduce the rDoF at a suitable temperature and rod length  \cite{Blackburn1996,AnoopRot}. To overcome this numerical difficulty and mimic the effect of very long rods that have vanishing rotational diffusivity, we manually freeze the rDoF by prohibiting the rod-shape impurities from rotating during the stress release in the minimization process (non-affine motion), although rotation is allowed during the affine transformation.

The same samples for different $c_d$ and for different $L_r$ used in the previous analysis (Fig.\ref{fig3} and Fig.\ref{fig4}) are taken and subjected to deformation with frozen rDoF. Fig.~\ref{fig5}(a) and Fig.~\ref{fig5}(c) show the stress-strain curves for the systems in two and three dimensions, with frozen rotational DoF of dimers. Firstly, the drastic increase in the yield strain shown in Fig.~\ref{fig3} has disappeared, reconfirming the vital role of the rotational degrees. Although $\gamma_Y$ increases slightly with increasing $c_d$ due to the higher stability of the system captured by $\Theta$ in Fig.~\ref{fig2}(d). Secondly, the maximum of $\chi_{dis}$ increases significantly along with a significant decrease in the width of $\chi_{dis}-\gamma$ curves shown in Fig.~\ref{fig5}(b) for 2D. The trend in 3D is even dramatic as shown in Fig.~\ref{fig5} (d), in which the stress-strain curve becomes nearly discontinuous at the transition point and $\chi_{dis}$ attains a very high peak value with a smaller spread in the loading axis. In Fig.~\ref{fig5}(e), the stress-strain curves are shown for various rod lengths $L_r$ for $c_r=0.1$. For all values of $L_r$ the failure is brittle, but longer rod impurities increase the shear modulus due to the higher stability of the samples (see Fig.~\ref{fig4}(e)), and cause the failure to become more brittle as implied from the peak height of $\chi_{dis}$ shown in panel (f).

Fig.~\ref{fig5}(g) shows the spatial variation of the non-affine displacement $D_{min}^2$, similar to Fig.~\ref{fig4}(f), with respect to the distance from the center of the shear band for both with and without rDoF. The extremely small displacement outside the shear band for the frozen rDoF scenario, along with increased displacement in the shear band region, demonstrates the highly brittle nature of the failure. This is also supported by the $D_{min}^2$ map in Fig.~\ref{fig5}(h).

\begin{figure}
\includegraphics[width=0.49\textwidth]{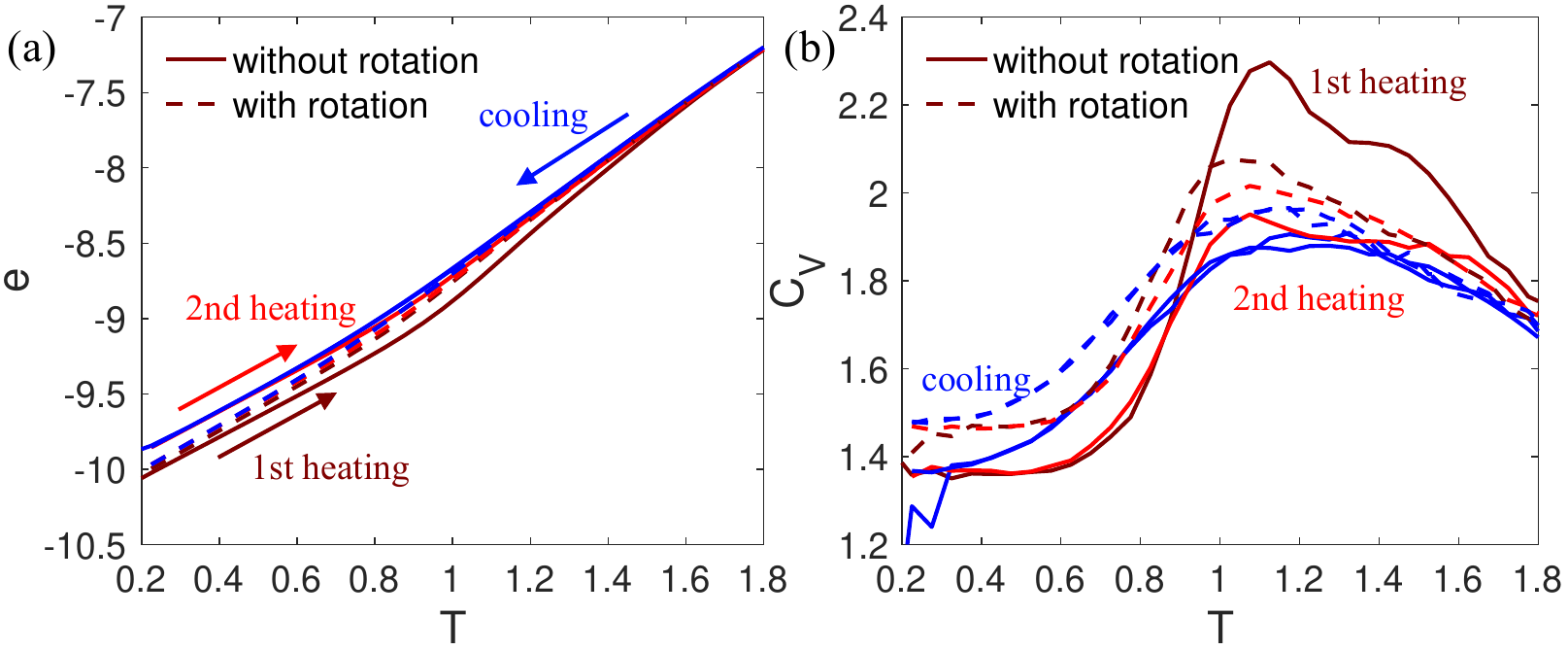}
\caption{\textbf{Ultra-stability with frozen rDoF: Kinetic aspect: } (a) The potential energy per particle $e$ is plotted for a system with $10\%$ dimers subjected to heating-cooling cycles at the rate of $dT/dt=10^{-4}$ (same as the rate with which the sample was prepared). The solid lines are for a system with frozen rotational DoFs; the observed hysteresis during the first cycle indicates the ultra-stable character. The second heating cycle does not show any hysteresis. The dashed lines are for the system evolved with rotational DoFs, and the absence of hysteresis is seen as expected. (b) The specific heat $C_V=de/dT$ is calculated from data in (a) by numerical differentiation and conveys the same}
\label{fig6}
\end{figure}

\vskip +0.1in
\noindent{\bf Kinetic aspect:}
After demonstrating their enhanced mechanical stability, we now focus on the kinetic stability of systems with frozen rDoF. To assess the kinetic stability of a solid, we observe the potential energy per particle (e) while subjecting it to heating-cooling cycles. An ultra-stable state would be characterized by a deep potential energy minimum that can not be reached through normal cooling, therefore resulting in hysteresis in the potential energy versus temperature plot. To test this, first, we prepare initial states of the system with dimer impurities using a cooling rate $\dot{T} = {10}^{-4}$. We choose $c_d = 0.1$ and $\sigma_r = 2.0$. During preparation, the dimer impurities are allowed to rotate. Next, we melt the system by heating it to $T = 1.8$ using the same $\dot{T}$ keeping the rDoF frozen, then cool it down to the same temperature using $\dot{T}$. Note that during the preparation of the sample, the rDoF are not kept frozen, but during the heating/cooling cycles, they are since we are interested in the gained kinetic stability due to the lack of rDoF. As shown in Fig.~\ref{fig6}(a), hysteresis is observed, validating the classical ultra-stable characteristics. Upon heating, the system remained in its glassy state until $T=1.12$, while on cooling, it reached its glassy state at $T=1.0$, resulting in a glass with a higher potential energy per particle. However, the second heat cycle did not exhibit hysteresis. Furthermore, the absence of hysteresis in the dashed lines of Fig.~\ref{fig6}(a), which depict the same procedure with rotationally free rods, confirms the effectiveness of rDoF in creating ultra-stable glasses. The specific heat ($C_V=de/dT$) plots in Fig.~\ref{fig6}(b) also support this conclusion.
\section*{Ductile-\lowercase{to}-Brittle transition with impurities}

\AM{Finally, we study the transition from ductile to brittle yielding as the concentration of rotationally frozen, rod-shaped impurities increases. Here, ductile yielding refers to a smooth stress-strain curve in the thermodynamic limit, while brittle yielding indicates a discontinuous stress-strain curve. As discussed previous studies \cite{RossiTarju2022,Ozawa2018,mutnejaBPB2025} have shown that these two distinct types of yielding can be exhibited by the same material, depending on the degree of annealing to which the material is subjected. The well-annealed, and thus more stable, sample undergoes brittle failure, whereas the less-annealed sample fails in a ductile manner. These two regimes are separated by a critical point associated with a finite critical inherent disorder strength. It was argued in these studies that the finite critical disorder above which the samples will go through a ductile yielding process survives in the thermodynamic limit. Various critical exponents obtained from finite-size analysis strongly suggest that this criticality belongs to the random-field Ising model (RFIM) universality class \cite{rossi2023far-from-equilibrium-5f0}. Recently, the brittle-to-ductile transition has also been demonstrated in Ref. \cite{mutnejaBPB2025} by varying the concentration of randomly pinned impurity particle, which introduces additional quenched disorder along with the system’s inherent disorder strength.  \Rev{Random pinning is not easily realizable in experiments, whereas in this work, we present a novel ductile-to-brittle transition, which is easily accessible in both molecular glass and colloidal glass experiments, that occurs due to an increase in structural amorphous order in the sample due to the addition of elongated rod-like impurities. This is illustrated in previous sections (Figs. \ref{fig2}(d) and \ref{fig4}(e)). The transition results from addition of rotationally frozen asymmetric impurities.  The same can be achieved by having elongated impurities with large aspect ratio but for computational difficulty we demonstrate this using rotationally frozen impurities with not so large aspect ratio.  As discussed below,  these results seem to indicate that the ductitle to brittle transition  also belongs to the same RFIM universality class. }}
\begin{figure}
\includegraphics[width=0.47\textwidth]{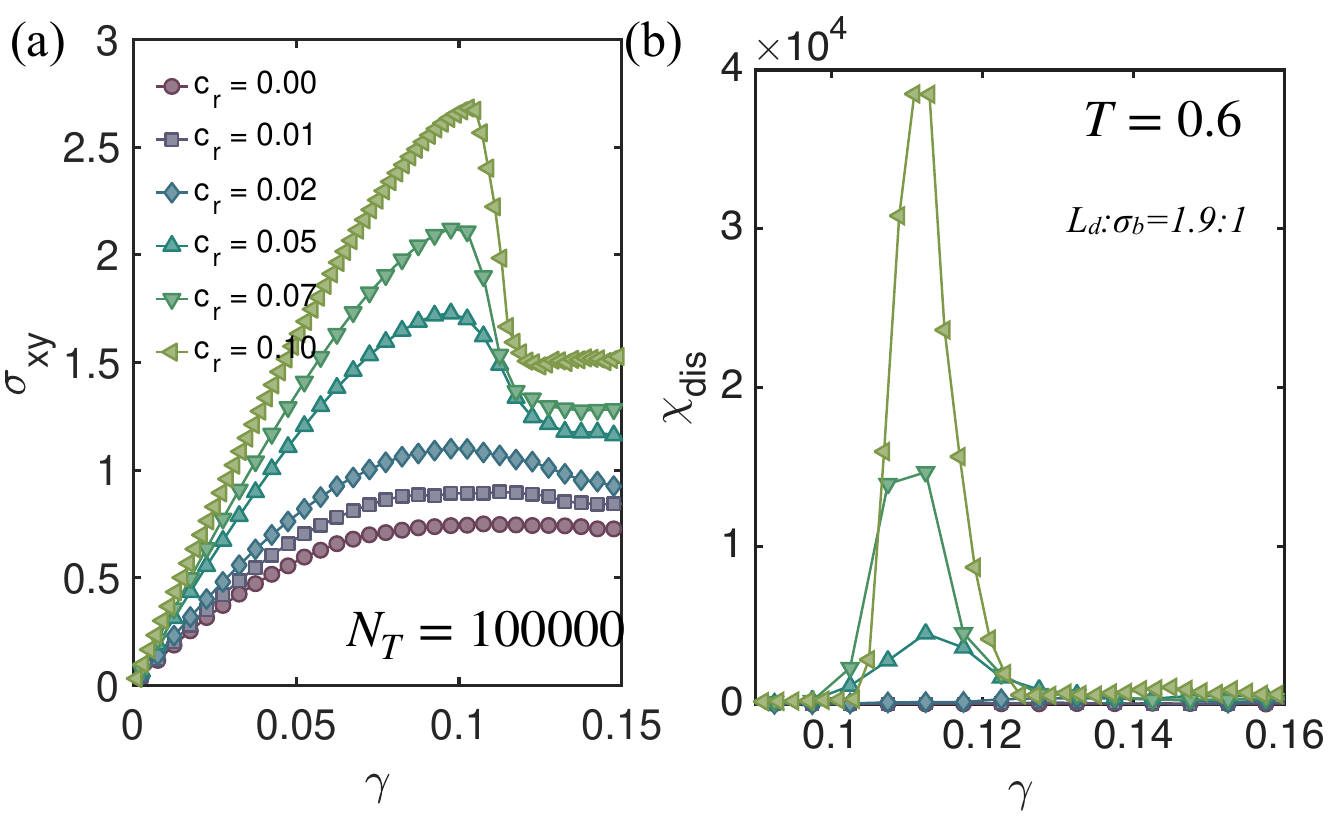}
\caption{\Rev{\textbf{Ductile-to-brittle transition with the addition of rotationally frozen rod impurities:}  (a) The averaged stress-strain response showing the ductile to brittle like response with increasing concentrations of rotationally frozen (frozen non-affine rotations) impurities for a poorly annealed sample,.  (b) Their corresponding
susceptibilities to show the strong growth of the peak of the susceptibility with increasing rod concentrations. }}
\label{SI4}
\end{figure}

\begin{figure*}
\includegraphics[width=0.97\textwidth]{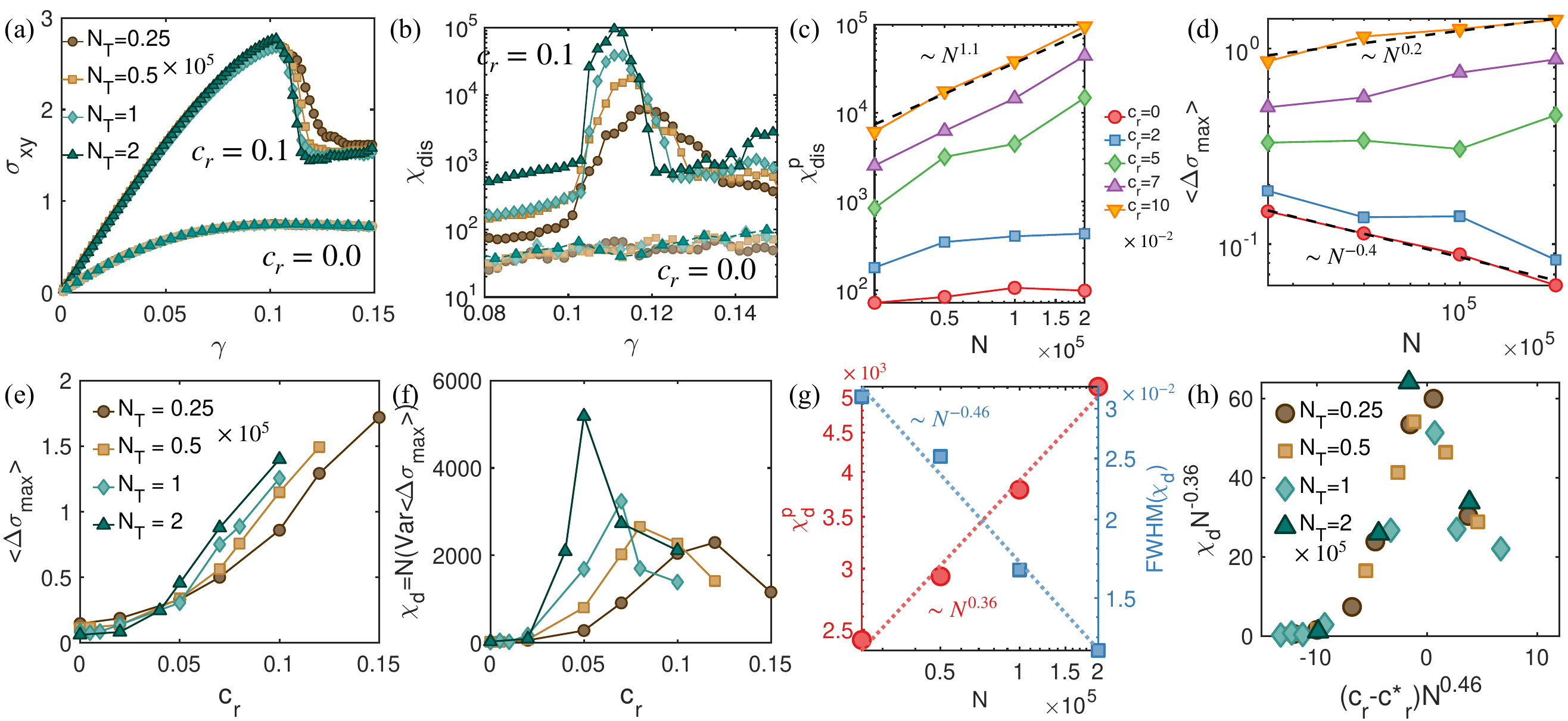}
\caption{\textbf{Impurity driven ductile-to-brittle transition:} (a) System size dependence of stress-strain curves for the 3dKA system prepared at a high temperature of $T = 0.6$. The pure system, shown in lighter colors, exhibits a ductile macroscopic response with no system size dependence, while systems with $10\%$ rod impurities (aspect ratio $1.9:1$) show progressively brittle behavior as system size increases. (b) Susceptibility plots for both the pure and doped systems at various system sizes. (c) The system size dependence of peak height for different impurity concentrations. For $c_r\le0.02$, the $\chi_{dis}^p$ saturates, while it follows the indicated power law for $c_r\ge0.05$. (d) The system size dependence of  $\langle\sigma_{\text{max}}\rangle$ with increasing $c_r$. The response shifts from vanishing to not-vanishing with increasing $c_r$ in the thermodynamic limit. (e) The increase in $\langle\sigma_{\text{max}}\rangle$ with increasing $c_r$ becomes sharper with increasing system size. (f) Ensemble-level fluctuations of $\langle\sigma_{\text{max}}\rangle$ reveal a nonmonotonic peak structure, marking the transition point. As system size increases, the peak narrows, and its height increases, highlighting the critical nature of the transition. (g) The peak height diverges as $\chi_d^p \sim N^{0.36}$, while the FWHM vanishes as $\text{FWHM} \sim N^{-0.46}$. (h) The obtained collapse of data points in panel (f) using the determined power laws.}
\label{fig7}
\end{figure*}

For an inherently ductile system, we choose an equilibrated state at relatively high temperature, $T = 0.6$, as a quiescent state, which, when prepared with \textit{highly-asymmetric} rod-impurities, undergoes brittle failure under AQS conditions with frozen rotational DoF.  \Rev{The stress-strain response for such systems undergoing ductile-to-brittle transition with increasing impurity concentration is shown in Fig.~\ref{SI4}(a), along the respective susceptibility in panel (b).  One can clearly see the ductile response of a poorly annealed glass changes systematically to brittle response with increasing concentration of rod impurities without any further temperature annealing. The rapid rise of the susceptibility peak corroborates this change over.  To systematically study the transition in} the thermodynamic, $N\rightarrow\infty$ limit, we performed detailed finite size scaling (FSS) analysis. The averaged stress response for pure and doped ($10\%$ rods of length $L_r = 1.9$) systems with varying system sizes is shown in Fig.~\ref{fig7}(a). It is evident that the pure system (lighter colors) remains ductile with no observable system size dependence, while the system with impurities progressively becomes more brittle as the system size increases (shown in darker colors). This transformation is further corroborated by the susceptibility plots in Fig.~\ref{fig7}(b): the susceptibility amplitude for the pure system (lighter colors) is small and non-responsive to system size (also see  Fig.~\ref{fig7}(c)), indicating ductile behavior. In stark contrast, the susceptibility peak for systems with $10\%$ impurities (darker colors) grows immensely, and the response sharpens with increasing system size, a typical feature of brittle yielding transition \cite{Ozawa2018}. In Fig.~\ref{fig7}(c), the susceptibility peak is shown to grow as a power law with an exponent of $\sim1.1$ for systems with $c_r\ge0.05$, while it saturates for $c_r\le0.02$, indicating the transition happening at impurity concentration between $0.02<c_r^*(N\rightarrow\infty)<0.05$., \Rev{which agrees well with the large $N$ estimates of the critical point of the transition as discussed below.}

Furthermore, the largest plastic drop throughout loading, $\Delta \sigma_{max}$, is usually taken as a good order parameter to characterise such ductile-to-brittle transition \cite{RossiTarju2022,Ozawa2018}. For ductile yielding,  $\langle\Delta \sigma_{max}\rangle$ vanishes in the thermodynamic limit, but it remains non-zero for brittle yielding. These two contrasting system size responses can be seen for $c_r = 0$ and  $c_r = 0.1$ in Fig.~\ref{fig7}(d). The averaged order parameter for $c_r = 0$ systems decays to zero following a power law of $\sim N^{-0.4}$, while it remains finite and even grows for $c_r = 0.1$ systems following  $\langle\Delta \sigma_{max}\rangle\sim N^{0.2}$, clearly pointing out the ductile-to-brittle transition with increasing $c_r$. The $c_r$ variation of the order parameter shown in Fig.~\ref{fig7}(e) also increases sharply for larger system sizes. 

Further,  to demonstrate the critical nature of the transition, we focus on the fluctuation of $\Delta \sigma_{max}$. In Fig.~\ref{fig7}(f), the susceptibility, given by the fluctuation of the order parameter, $\chi_d = N\left( \langle \Delta \sigma_{max}^2 \rangle - \langle \Delta \sigma_{max} \rangle^2\right)$, is plotted as a function of $c_r$ for various system sizes. $\chi_d$ exhibits non-monotonic dependence on $c_r$, with a sharp peak at $c_r^{*}(N)$, which becomes sharper and attains larger peak amplitude with increasing system size, confirming the critical nature. Fig.~\ref{fig7}(g) shows that the peak height diverges following $\chi_d^p \sim N^{0.36}$ as $N \rightarrow \infty$, and the full width at half maximum (FWHM) of the $\chi_d - c_r$ curve vanishes following $\text{FWHM} \sim N^{-0.46}$ as $N \rightarrow \infty$. These power laws are then used to obtain a $\chi_d - c_r$ data collapse as shown in Fig.~\ref{fig7}(h) for different system sizes. The quality of collapse and thermodynamic divergence of fluctuation at a critical impurity concentration clearly indicates the \Rev{existence of a ductile-brittle non-equilibrium transition at a finite critical disorder strength, in complete agreement with the results in \cite{RossiTarju2022,Ozawa2018,mutnejaBPB2025}, in which the disorder is varied by completely different way. }

\begin{figure}
\includegraphics[width=0.49\textwidth]{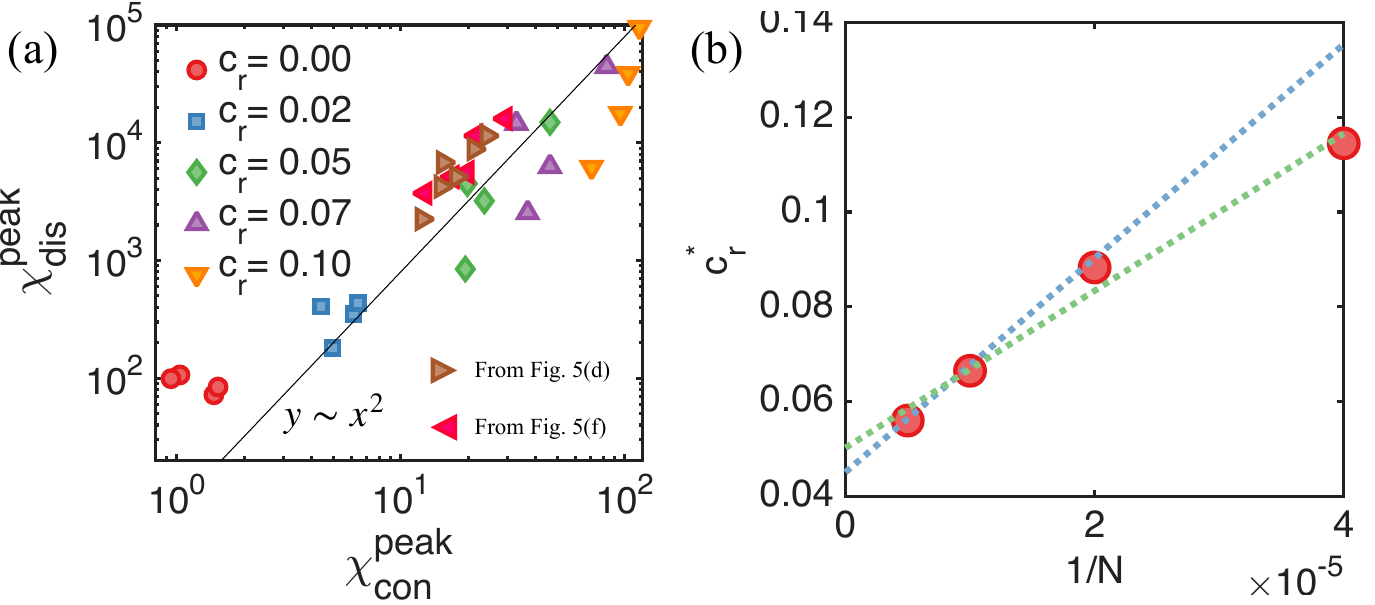}
\vskip -0.1in
\caption{\Rev{\textbf{Finite Size Scaling of Critical Concentration:} (a) Scatter plot of peak height of disconnected part of the susceptibility $\chi_{dis}^{p}$ as a function of peak height of the connected part of the susceptibility, $\chi_{con}^p$. RFIM universality class indicates a quadratic relation between them as $\chi_{dis} \sim \chi_{con}^2$, which is exact in the mean field limit. The data are in good agreement with that predictions. (b) Critical impurity concentration, $c_r^*(N)$ as a function of inverse system size to find the value of the critical concentration in the thermodynamic limit. Straight line fits with first $3$ points gives $c_r^*(N\to\infty) \simeq 0.045$ as compared to $c_r^*(N\to\infty) \simeq 0.050$ if all the $4$ data points are included in the fitting. This strongly indicates a finite disorder critical point in Brittle-Ductile transition at thermodynamically large system sizes.}}
\label{rfim_fss}
\end{figure}

\Rev{As suggested in Ref.\cite{Ozawa2018}, a strong evidence of the underlying critical point, the first order phase transition between a Ductile and Brittle state and the possibility of the corresponding university class of the transition being the Random Field Ising Model (RFIM) can be ascertained by studying the two different susceptibilities: the connected susceptibility $\chi_{con} = -d\langle\sigma\rangle/d\gamma$ and the disconnected one $\chi_{dis}$ \ref{eqn:eqn1}. It is argued in \cite{Ozawa2018} that a strong signature of existence of random field like disorder in the system can be probed by studying the mutual relationship between $\chi_{dis}$ and $\chi_{con}$ and random field like disorder will predict $\chi_{dis} \sim \chi_{con}^2$ in the mean field limit. This relationship is also found to hold at finite dimensions near a first order phase transition. In Fig.\ref{rfim_fss}(a), we have plotted the peak values of these two different susceptibilities and within our studied range, the relation is obeyed approximately.  We also found that peak of $\chi_{con}$ grows with system size as $\chi_{con} \sim N^{1/2}$ for impurity concentration above the critical concentration,  $c_r > c_r^*$.  On the other hand,  the peak show almost no size effect or decreases with increasing system size mildly for $c_r < c_r^*$ (see SI Fig.4).  These results are in close agreement with the results reported in \cite{RossiTarju2022,Ozawa2018}. This suggests that the ductile-brittle transition induced in our system via elongated impurities might as well belong to the same universality class of RFIM. Note that in previous studies the transition is studied by thermal annealing of the sample whereas in this study the transition is tuned by adding rod-like impurities and brittle state appears at larger concentration of the impurities which is a non-trivial effect. } 

\Rev{Rod-like impurities in our samples strengthen the solids in a manner similar to effective annealing, rather than making them more disordered and ductile. Additionally, doping these elongated impurities leads to significant strengthening of the materials, as evidenced by the values of the structural order shown in Fig. \ref{fig4}(e). The $\Theta$ order parameter decreases significantly with an increasing aspect ratio of the impurity, more so than what is achieved by the slowest thermal annealing rate studied in this work.  It is important to highlight that our proposed method for accessing the brittle-ductile transition using elongated rod-like impurities is much more experimentally feasible than achieving the same effect through thermal annealing, particularly for soft materials such as colloidal glasses.  Next, we examine the system size dependence of the critical disorder strength as a function of system size \( N \). In panel (b) of the same figure, we present the critical disorder strength as a function of \( 1/N \) and fit the data with a straight line to determine the infinite system size limit. By fitting the first three points with a straight line, we found the thermodynamic critical disorder \( c_r^*(N \to \infty) \simeq 0.045 \) based on the intercept at zero. Conversely, fitting all four data points yields a value of \( 0.05 \). This result suggests that the extrapolated thermodynamic limit of the critical disorder is relatively large and finite, which contrasts with the notion that no such transition exists at any disorder strength.}

\Rev{The investigation into the critical-like behavior of the brittle-ductile transition in systems with finite critical disorder presents exciting opportunities for understanding the mechanisms that influence material strength.  In systems where the disorder strength can be manipulated through methods such as temperature annealing, random particle pinning, or micro-alloying with rod-shaped particles, there is a compelling case for a thermodynamic non-equilibrium transition from ductile to brittle states. TThis transition appears to fall within the universality class of the random-field Ising model (RFIM), as noted also in \cite{RossiTarju2022}.  Combining thermal annealing through swap Monte Carlo methods with doping using rod-like impurities could provide valuable insights into this transition over a wider range of annealing conditions. Additionally, it is crucial to extend the analysis to a broader variety of system sizes. This expansion will not only enhance our understanding of the transition but will also be essential for developing accurate macroscopic descriptions of plastic behavior in amorphous solids, which are important for various industrial applications.}

\section*{Discussion}
Our findings reveal that the addition of impurities of larger size into a glassy system can significantly increase its ability to withstand external loads beyond its usual limit, along with increased shear modulus and yield stress. Furthermore, rod-shaped asymmetric inclusions with an aspect ratio close to unity, which have both translational and rotational degrees of freedom, are much better for microalloying, as they can significantly increase the yield strain, yield stress, and shear modulus even at small concentrations. Conversely, inclusions with frozen or constrained rDoF, which can be achieved by increasing the rod length while keeping the diameter the same, exhibit comparatively mild enhancement of the yield strain and vanishing plasticity at small strain, but exhibit more brittle-like catastrophic failure, suggesting the formation of ultra-stable samples. Recently, in Ref.\cite{Ozawa2023}, the bulk ultra-stable glass phase was claimed to be formed by randomly bonding the nearest neighbors of an otherwise poorly annealed glassy state. Such bonded molecules would have significantly less rotational freedom because of the packing. Thus, our results offer an alternate explanation of the results reported in Ref.\cite{Ozawa2023}. 

\Rev{Most importantly we show a sharp ductile-to-brittle transition induced in a poorly annealed glass sample with rotationally frozen impurities and demonstrate the existence of a critical point at a critical $c_r^*$ that separates ductile and brittle yielding.  Systems doped with rod-like impurities above the critical concentration will have lower disorder (lower than critical), and thus will fail catastrophically via brittle yielding. Observation of a non-equilibrium transition between brittle and ductile states using rod-like impurities opens up the possibility of studying this transition in experiments via micro-alloying techniques, both in molecular glasses and colloidal glasses. Especially, in soft glasses like colloidal glasses where achieving higher degree of annealing via densification might not be feasible, aspherical impurities will probably be the only viable option for tuning their mechanical properties as well as driving the system through a ductile to brittle transition. Although the results align with the ideas of previous studies, the microscopic understanding of the ductile-to-brittle transition with increasing concentration of rotationally constrained impurities remains unclear and is an interesting topic for future investigation.}

The observation of the long or rotationally frozen rod-shaped impurities making the system behave in a highly brittle way can be hypothesized as an effect of the larger length scale in the system. The ultra-stable states are sampled from extremely low temperatures, implying a sizable structural length scale, while the states sampled from high temperatures would have a structural length scale of a few particle diameters. By inserting a rod of length $L_r$, a static correlation of the same length is induced. Thus, the observed similarity between systems doped with larger rod-like impurities and ultra-stable glasses may simply be due to the increased static correlation length. Although we do not have direct proof of this argument, it seems more likely to be the scenario, and further work is needed to understand the microscopic reason for the strong similarity between ultra-stable glasses and microalloyed glasses with long rod-like impurities. \Rev{It will be very interesting to study dynamical properties of these microalloyed glasses in their supercooled liquid regime and compute the growth of point-to-set (PTS) correlation length and compare that with the correlation length obtained in ultra-stable glasses. }

Finally,  it is important to highlight that the incorporation of aspherical impurities into realistic amorphous solids is not a difficult process. This approach enables one to adjust the mechanical properties of these disordered materials by varying the aspect ratio of the impurities. Spherically symmetric impurities do not introduce additional rotational degrees of freedom (rDoF); however, the inclusion of slightly asymmetric impurities, which are equal to or larger than the constituent particles, results in a substantial increase in both the yield strain and yield stress of the material due to the introduction of rDoF. Conversely, the addition of highly aspherical, rod-like impurities may effectively convert the system into one that more closely resembles an ultra-stable configuration.  \Rev{Consequently, our findings present a systematic and controlled methodology for microalloying that can be readily implemented in both molecular glasses and soft glasses for prospective applications.}

\section*{Methods}{\label{methods}
We conduct simulations of a binary Kob-Anderson mixture of Lennard-Jones particles in both two and three dimensions. The model details are as follows:
\vskip +0.1in
\noindent{\bf \large 3d model:} The 3D Kob-Anderson model (referred to as 3D) \cite{KA} represents a binary mixture comprising A- and B-type Lennard-Jones particles, with a concentration ratio of $80 : 20$. This generic model resembles molecular glass-forming liquid, $Ni80P20$. The following potential governs the interaction between particles:
\begin{equation}
V_{\alpha\beta}(\textbf{r})=4\epsilon_{\alpha\beta}
\left[\left(\frac{\sigma_{\alpha\beta}}{r}\right)^{12}-\left(\frac{\sigma
_{\alpha\beta}}{r}\right)^6\right]
\label{eqn:eqn3}
\end{equation}
where $\alpha$ and $\beta$ vary in A, B and the interaction strengths and radii are $\epsilon_{_{AA}}=1.0$, $\epsilon_{_{AB}}=1.5$, $\epsilon_{_{BB}}=0.5$; $\sigma_{_{AA}}=1.0$, $\sigma_{_{BB}}=0.88$ and $\sigma_{_{AB}}=0.8$. The interaction is truncated at $r=2.5\sigma_{\alpha\beta}$ and is smoothed by adding up to $2^{nd}$ order terms.

\vskip +0.1in
\noindent{\bf \large 2d model:} 
2d modified Kob-Anderson model (mentioned as 2d)  \cite{2dmKA} is the glass forming model in 2 dimensions with properties like 3dKA. It is a 65:35 binary mixture of the same A and B particles of the 3dKA model interacting with the same potential and parameters.

\vskip +0.1in
\noindent{\bf \large Rods and spherical impurities:}
In the glass formers mentioned above, we added $c_{r}$ ($c_{s}$) concentration of rods (spherical particles) in the parent system of $N=(1-c_{x})N_T$ particles; we used a system with $N_T=100000$ total particles. Each rod is formed by attaching $nb$ spheres at a fixed distance of $d=0.3\sigma_{AA}$.
In this study, to achieve the soft pinning effect, the rods are made up of two beads each with diameter of $\sigma_{b}=2.0$, while $\sigma_{b,\alpha}=0.5(\sigma_{b}+\sigma_{\alpha})$, same is true for the spherical impurities; while they have same mass and interacts via same potential as parent spheres with $\epsilon_{b,\alpha}=1$ and $\epsilon_{b,b}=\frac{1}{2}$ for both the models. $\epsilon_{b,b}=\frac{1}{2}$ is chosen to avoid the nematic ordering. The aspect ratio of such a dimer will be $L_r:\sigma_b=2.3:2$. We have also studied the mechanical response with thin rod inclusions with $\sigma_{b}=1.0$. The aspect ratio of such a impurity will be $L_r:\sigma_b=(n_b-1)*0.3+1:1$.

\vskip +0.1in
\noindent{\bf \large Sample preparation:}
The moderately-annealed state of a glassy system with  $c_{r}$ ($c_s$), the concentration of rods (spherical impurities ) is prepared by firstly equilibrating the system at high temperature ($T=0.5$). It is then slowly cooled to temperature $T=5\times10^{-4}$, with a cooling rate of  $dT/dt=1\times10^{-4}$. This annealed state is minimized via conjugate-gradient to reach the inherent state (IS). For high temperature studies, the sample is simply equilibrated at $T=0.60$, and then quenched to the IS. This inherent state is then used as a starting point for all shear procedures.

\vskip +0.1in
\noindent{\bf \large  Shear protocol:}
This work focuses only on the athermal quasi-static ($T \to 0$ and $\dot{\gamma} \to 0$) deformation. We start with an inherent state and deform it by increasing the strain by  $\delta \gamma = 5\times10^{-5}$ in every step. Each deformation step contains two parts; the first step is called affine transformation, in which particle positions are modified in the following way: $x_i = x_i + \delta \gamma y_i$, $y_i = y_i$, $z_i = z_i$. Here the strain is applied in the $x$ direction. The second step involves minimization of energy in which particles are brought back to mechanical equilibrium. We use the conjugate gradient method for energy minimization.

\vskip +0.1in
\noindent{\bf \large  Structural order parameter:}
The structural order parameter utilized in this study aims to characterize the deviation of local structure from perfect steric packing, as outlined in Ref.\cite{Tong2019}. The specific order parameter, denoted as $\Theta_o$ for a tagged particle $o$, is determined through a series of steps detailed next. Firstly, the neighbors of $o$ are identified using radical Voronoi tessellation, then sets of four particles, comprising the tagged particle $o$ and three of its neighbors, are identified to form tetrahedra. For each bond within the tetrahedron, the imperfection is calculated as $\sum_{<ab>} |r_{ab} - \sigma_{ab}|$, where $<ab>$ ranges over the six edges of the tetrahedron, and $\sigma_{ab} = (\sigma_a + \sigma_b) \times 0.5$ represents the favored distance. Normalization is applied, and the imperfections are averaged over all tetrahedra. The order parameter for a particle is then given by the expression:
\begin{equation}\label{eqn:eqn4}
\Theta_o=\frac{1}{N_0^{tetra}}\sum_{oijk}\frac{\sum_{<ab>}|r_{ab}-\sigma_{ab}|}{\sum_{<ab>}\sigma_{ab}}
\end{equation}
where $<oijk>$ runs over all tetrahedron sets. The system's structural order parameter is computed by averaging over all particles. A higher value of $\Theta_o$ indicates a more disordered local surrounding; thus, the $\Theta$ map serves as a valuable tool for identifying regions with shear bands, as elaborated in the results section. 

For systems with impurities, to directly compare structural order parameters, $\Theta_o$ is averaged only over the parent liquid particles. Additionally, for $\Theta_o$ calculation, only tetrahedra containing parent particles are considered, while Voronoi tessellation is performed on all particles, including the beads of each rod. We use open source software Voro++ \cite{PhysRevE.74.021306} to perform all Voronoi tessellations. 
}


\noindent{\bf \large Acknowledgement:}
SK acknowledges funding by intramural funds at TIFR Hyderabad from the Department of Atomic Energy (DAE) under Project Identification No. RTI 4007. Core Research Grant CRG/2019/005373 and Swarna Jayanti Fellowship SB/SFJ/2019-20/05 from Science and Engineering Research Board (SERB) are acknowledged for generous funding. Most of the computations are done using the HPC clusters bought using CRG/2019/005373 grant and Swarna Jayanti Fellowship, grants DST/SJF/PSA01/2018-19, and SB/SFJ/2019-20/05 of SK.



\bibliography{Citations}

\begin{thebibliography}{61}%
\makeatletter
\providecommand \@ifxundefined [1]{%
 \@ifx{#1\undefined}
}%
\providecommand \@ifnum [1]{%
 \ifnum #1\expandafter \@firstoftwo
 \else \expandafter \@secondoftwo
 \fi
}%
\providecommand \@ifx [1]{%
 \ifx #1\expandafter \@firstoftwo
 \else \expandafter \@secondoftwo
 \fi
}%
\providecommand \natexlab [1]{#1}%
\providecommand \enquote  [1]{``#1''}%
\providecommand \bibnamefont  [1]{#1}%
\providecommand \bibfnamefont [1]{#1}%
\providecommand \citenamefont [1]{#1}%
\providecommand \href@noop [0]{\@secondoftwo}%
\providecommand \href [0]{\begingroup \@sanitize@url \@href}%
\providecommand \@href[1]{\@@startlink{#1}\@@href}%
\providecommand \@@href[1]{\endgroup#1\@@endlink}%
\providecommand \@sanitize@url [0]{\catcode `\\12\catcode `\$12\catcode
  `\&12\catcode `\#12\catcode `\^12\catcode `\_12\catcode `\%12\relax}%
\providecommand \@@startlink[1]{}%
\providecommand \@@endlink[0]{}%
\providecommand \url  [0]{\begingroup\@sanitize@url \@url }%
\providecommand \@url [1]{\endgroup\@href {#1}{\urlprefix }}%
\providecommand \urlprefix  [0]{URL }%
\providecommand \Eprint [0]{\href }%
\providecommand \doibase [0]{https://doi.org/}%
\providecommand \selectlanguage [0]{\@gobble}%
\providecommand \bibinfo  [0]{\@secondoftwo}%
\providecommand \bibfield  [0]{\@secondoftwo}%
\providecommand \translation [1]{[#1]}%
\providecommand \BibitemOpen [0]{}%
\providecommand \bibitemStop [0]{}%
\providecommand \bibitemNoStop [0]{.\EOS\space}%
\providecommand \EOS [0]{\spacefactor3000\relax}%
\providecommand \BibitemShut  [1]{\csname bibitem#1\endcsname}%
\let\auto@bib@innerbib\@empty
\bibitem [{\citenamefont {Barrat}\ and\ \citenamefont
  {Lemaltre}(2011)}]{Barrat2011}%
  \BibitemOpen
  \bibfield  {author} {\bibinfo {author} {\bibfnamefont {J.-L.}\ \bibnamefont
  {Barrat}}\ and\ \bibinfo {author} {\bibfnamefont {A.}~\bibnamefont
  {Lemaltre}},\ }\bibfield  {title} {\bibinfo {title} {Heterogeneities in
  amorphous systems under shear},\ }in\ \href
  {https://doi.org/10.1093/acprof:oso/9780199691470.003.0008} {\emph {\bibinfo
  {booktitle} {Dynamical Heterogeneities in Glasses, Colloids, and Granular
  Media}}}\ (\bibinfo  {publisher} {Oxford University Press},\ \bibinfo {year}
  {2011})\ pp.\ \bibinfo {pages} {264--297}\BibitemShut {NoStop}%
\bibitem [{\citenamefont {Nicolas}\ \emph {et~al.}(2018)\citenamefont
  {Nicolas}, \citenamefont {Ferrero}, \citenamefont {Martens},\ and\
  \citenamefont {Barrat}}]{Nicolas2018}%
  \BibitemOpen
  \bibfield  {author} {\bibinfo {author} {\bibfnamefont {A.}~\bibnamefont
  {Nicolas}}, \bibinfo {author} {\bibfnamefont {E.~E.}\ \bibnamefont
  {Ferrero}}, \bibinfo {author} {\bibfnamefont {K.}~\bibnamefont {Martens}},\
  and\ \bibinfo {author} {\bibfnamefont {J.-L.}\ \bibnamefont {Barrat}},\
  }\bibfield  {title} {\bibinfo {title} {Deformation and flow of amorphous
  solids: Insights from elastoplastic models},\ }\href
  {https://doi.org/10.1103/RevModPhys.90.045006} {\bibfield  {journal}
  {\bibinfo  {journal} {Rev. Mod. Phys.}\ }\textbf {\bibinfo {volume} {90}},\
  \bibinfo {pages} {045006} (\bibinfo {year} {2018})}\BibitemShut {NoStop}%
\bibitem [{\citenamefont {Bonn}\ \emph {et~al.}(2017)\citenamefont {Bonn},
  \citenamefont {Denn}, \citenamefont {Berthier}, \citenamefont {Divoux},\ and\
  \citenamefont {Manneville}}]{Bonn2017}%
  \BibitemOpen
  \bibfield  {author} {\bibinfo {author} {\bibfnamefont {D.}~\bibnamefont
  {Bonn}}, \bibinfo {author} {\bibfnamefont {M.~M.}\ \bibnamefont {Denn}},
  \bibinfo {author} {\bibfnamefont {L.}~\bibnamefont {Berthier}}, \bibinfo
  {author} {\bibfnamefont {T.}~\bibnamefont {Divoux}},\ and\ \bibinfo {author}
  {\bibfnamefont {S.}~\bibnamefont {Manneville}},\ }\bibfield  {title}
  {\bibinfo {title} {Yield stress materials in soft condensed matter},\ }\href
  {https://doi.org/10.1103/RevModPhys.89.035005} {\bibfield  {journal}
  {\bibinfo  {journal} {Rev. Mod. Phys.}\ }\textbf {\bibinfo {volume} {89}},\
  \bibinfo {pages} {035005} (\bibinfo {year} {2017})}\BibitemShut {NoStop}%
\bibitem [{\citenamefont {Ghosh}\ and\ \citenamefont
  {Schweizer}(2023)}]{GhoshJOR2023}%
  \BibitemOpen
  \bibfield  {author} {\bibinfo {author} {\bibfnamefont {A.}~\bibnamefont
  {Ghosh}}\ and\ \bibinfo {author} {\bibfnamefont {K.~S.}\ \bibnamefont
  {Schweizer}},\ }\bibfield  {title} {\bibinfo {title} {{Microscopic activated
  dynamics theory of the shear rheology and stress overshoot in ultradense
  glass-forming fluids and colloidal suspensions}},\ }\href
  {https://doi.org/10.1122/8.0000546} {\bibfield  {journal} {\bibinfo
  {journal} {Journal of Rheology}\ }\textbf {\bibinfo {volume} {67}},\ \bibinfo
  {pages} {559} (\bibinfo {year} {2023})},\ \Eprint
  {https://arxiv.org/abs/https://pubs.aip.org/sor/jor/article-pdf/67/2/559/19825696/559\_1\_online.pdf}
  {https://pubs.aip.org/sor/jor/article-pdf/67/2/559/19825696/559\_1\_online.pdf}
  \BibitemShut {NoStop}%
\bibitem [{\citenamefont {Hentschel}\ \emph {et~al.}(2011)\citenamefont
  {Hentschel}, \citenamefont {Karmakar}, \citenamefont {Lerner},\ and\
  \citenamefont {Procaccia}}]{HKLPPRE2011}%
  \BibitemOpen
  \bibfield  {author} {\bibinfo {author} {\bibfnamefont {H.~G.~E.}\
  \bibnamefont {Hentschel}}, \bibinfo {author} {\bibfnamefont {S.}~\bibnamefont
  {Karmakar}}, \bibinfo {author} {\bibfnamefont {E.}~\bibnamefont {Lerner}},\
  and\ \bibinfo {author} {\bibfnamefont {I.}~\bibnamefont {Procaccia}},\
  }\bibfield  {title} {\bibinfo {title} {Do athermal amorphous solids exist?},\
  }\href {https://doi.org/10.1103/PhysRevE.83.061101} {\bibfield  {journal}
  {\bibinfo  {journal} {Phys. Rev. E}\ }\textbf {\bibinfo {volume} {83}},\
  \bibinfo {pages} {061101} (\bibinfo {year} {2011})}\BibitemShut {NoStop}%
\bibitem [{\citenamefont {Maloney}\ and\ \citenamefont
  {Lema\^{\i}tre}(2006)}]{MaloneyLemaitrePRE2006}%
  \BibitemOpen
  \bibfield  {author} {\bibinfo {author} {\bibfnamefont {C.~E.}\ \bibnamefont
  {Maloney}}\ and\ \bibinfo {author} {\bibfnamefont {A.}~\bibnamefont
  {Lema\^{\i}tre}},\ }\bibfield  {title} {\bibinfo {title} {Amorphous systems
  in athermal, quasistatic shear},\ }\href
  {https://doi.org/10.1103/PhysRevE.74.016118} {\bibfield  {journal} {\bibinfo
  {journal} {Phys. Rev. E}\ }\textbf {\bibinfo {volume} {74}},\ \bibinfo
  {pages} {016118} (\bibinfo {year} {2006})}\BibitemShut {NoStop}%
\bibitem [{\citenamefont {Maloney}\ and\ \citenamefont
  {Lema\^{\i}tre}(2004)}]{MaloneyLemaitrePRL2004}%
  \BibitemOpen
  \bibfield  {author} {\bibinfo {author} {\bibfnamefont {C.}~\bibnamefont
  {Maloney}}\ and\ \bibinfo {author} {\bibfnamefont {A.}~\bibnamefont
  {Lema\^{\i}tre}},\ }\bibfield  {title} {\bibinfo {title} {Subextensive
  scaling in the athermal, quasistatic limit of amorphous matter in plastic
  shear flow},\ }\href {https://doi.org/10.1103/PhysRevLett.93.016001}
  {\bibfield  {journal} {\bibinfo  {journal} {Phys. Rev. Lett.}\ }\textbf
  {\bibinfo {volume} {93}},\ \bibinfo {pages} {016001} (\bibinfo {year}
  {2004})}\BibitemShut {NoStop}%
\bibitem [{\citenamefont {Karmakar}\ \emph {et~al.}(2010)\citenamefont
  {Karmakar}, \citenamefont {Lerner}, \citenamefont {Procaccia},\ and\
  \citenamefont {Zylberg}}]{SmarajitLernerItamarJacquesPRE}%
  \BibitemOpen
  \bibfield  {author} {\bibinfo {author} {\bibfnamefont {S.}~\bibnamefont
  {Karmakar}}, \bibinfo {author} {\bibfnamefont {E.}~\bibnamefont {Lerner}},
  \bibinfo {author} {\bibfnamefont {I.}~\bibnamefont {Procaccia}},\ and\
  \bibinfo {author} {\bibfnamefont {J.}~\bibnamefont {Zylberg}},\ }\bibfield
  {title} {\bibinfo {title} {Statistical physics of elastoplastic steady states
  in amorphous solids: Finite temperatures and strain rates},\ }\href
  {https://doi.org/10.1103/PhysRevE.82.031301} {\bibfield  {journal} {\bibinfo
  {journal} {Phys. Rev. E}\ }\textbf {\bibinfo {volume} {82}},\ \bibinfo
  {pages} {031301} (\bibinfo {year} {2010})}\BibitemShut {NoStop}%
\bibitem [{\citenamefont {Mutneja}\ and\ \citenamefont
  {Schweizer}(2025)}]{mutneja2025}%
  \BibitemOpen
  \bibfield  {author} {\bibinfo {author} {\bibfnamefont {A.}~\bibnamefont
  {Mutneja}}\ and\ \bibinfo {author} {\bibfnamefont {K.~S.}\ \bibnamefont
  {Schweizer}},\ }\href {https://arxiv.org/abs/2503.07436} {\bibinfo {title}
  {Microscopic theory of nonlinear rheology and double yielding in dense
  attractive glass forming colloidal suspensions}} (\bibinfo {year} {2025}),\
  \Eprint {https://arxiv.org/abs/2503.07436} {arXiv:2503.07436 [cond-mat.soft]}
  \BibitemShut {NoStop}%
\bibitem [{\citenamefont {Eshelby}\ and\ \citenamefont
  {Peierls}(1957)}]{Eshelby1957}%
  \BibitemOpen
  \bibfield  {author} {\bibinfo {author} {\bibfnamefont {J.~D.}\ \bibnamefont
  {Eshelby}}\ and\ \bibinfo {author} {\bibfnamefont {R.~E.}\ \bibnamefont
  {Peierls}},\ }\bibfield  {title} {\bibinfo {title} {The determination of the
  elastic field of an ellipsoidal inclusion, and related problems},\ }\href
  {https://doi.org/10.1098/rspa.1957.0133} {\bibfield  {journal} {\bibinfo
  {journal} {Proceedings of the Royal Society of London. Series A. Mathematical
  and Physical Sciences}\ }\textbf {\bibinfo {volume} {241}},\ \bibinfo {pages}
  {376} (\bibinfo {year} {1957})}\BibitemShut {NoStop}%
\bibitem [{\citenamefont {Nandi}\ \emph {et~al.}(2016)\citenamefont {Nandi},
  \citenamefont {Biroli},\ and\ \citenamefont {Tarjus}}]{Nandi2016}%
  \BibitemOpen
  \bibfield  {author} {\bibinfo {author} {\bibfnamefont {S.~K.}\ \bibnamefont
  {Nandi}}, \bibinfo {author} {\bibfnamefont {G.}~\bibnamefont {Biroli}},\ and\
  \bibinfo {author} {\bibfnamefont {G.}~\bibnamefont {Tarjus}},\ }\bibfield
  {title} {\bibinfo {title} {Spinodals with disorder: From avalanches in random
  magnets to glassy dynamics},\ }\href
  {https://doi.org/10.1103/PhysRevLett.116.145701} {\bibfield  {journal}
  {\bibinfo  {journal} {Phys. Rev. Lett.}\ }\textbf {\bibinfo {volume} {116}},\
  \bibinfo {pages} {145701} (\bibinfo {year} {2016})}\BibitemShut {NoStop}%
\bibitem [{\citenamefont {Ozawa}\ \emph {et~al.}(2018)\citenamefont {Ozawa},
  \citenamefont {Berthier}, \citenamefont {Biroli}, \citenamefont {Rosso},\
  and\ \citenamefont {Tarjus}}]{Ozawa2018}%
  \BibitemOpen
  \bibfield  {author} {\bibinfo {author} {\bibfnamefont {M.}~\bibnamefont
  {Ozawa}}, \bibinfo {author} {\bibfnamefont {L.}~\bibnamefont {Berthier}},
  \bibinfo {author} {\bibfnamefont {G.}~\bibnamefont {Biroli}}, \bibinfo
  {author} {\bibfnamefont {A.}~\bibnamefont {Rosso}},\ and\ \bibinfo {author}
  {\bibfnamefont {G.}~\bibnamefont {Tarjus}},\ }\bibfield  {title} {\bibinfo
  {title} {Random critical point separates brittle and ductile yielding
  transitions in amorphous materials},\ }\href
  {https://doi.org/10.1073/pnas.1806156115} {\bibfield  {journal} {\bibinfo
  {journal} {Proceedings of the National Academy of Sciences}\ }\textbf
  {\bibinfo {volume} {115}},\ \bibinfo {pages} {6656} (\bibinfo {year}
  {2018})}\BibitemShut {NoStop}%
\bibitem [{\citenamefont {Popovi\ifmmode~\acute{c}\else \'{c}\fi{}}\ \emph
  {et~al.}(2018)\citenamefont {Popovi\ifmmode~\acute{c}\else \'{c}\fi{}},
  \citenamefont {de~Geus},\ and\ \citenamefont
  {Wyart}}]{WyartPRENonEqPhaseTran}%
  \BibitemOpen
  \bibfield  {author} {\bibinfo {author} {\bibfnamefont {M.}~\bibnamefont
  {Popovi\ifmmode~\acute{c}\else \'{c}\fi{}}}, \bibinfo {author} {\bibfnamefont
  {T.~W.~J.}\ \bibnamefont {de~Geus}},\ and\ \bibinfo {author} {\bibfnamefont
  {M.}~\bibnamefont {Wyart}},\ }\bibfield  {title} {\bibinfo {title}
  {Elastoplastic description of sudden failure in athermal amorphous materials
  during quasistatic loading},\ }\href
  {https://doi.org/10.1103/PhysRevE.98.040901} {\bibfield  {journal} {\bibinfo
  {journal} {Phys. Rev. E}\ }\textbf {\bibinfo {volume} {98}},\ \bibinfo
  {pages} {040901} (\bibinfo {year} {2018})}\BibitemShut {NoStop}%
\bibitem [{\citenamefont {Jaiswal}\ \emph {et~al.}(2016)\citenamefont
  {Jaiswal}, \citenamefont {Procaccia}, \citenamefont {Rainone},\ and\
  \citenamefont {Singh}}]{Jaiswal2016}%
  \BibitemOpen
  \bibfield  {author} {\bibinfo {author} {\bibfnamefont {P.~K.}\ \bibnamefont
  {Jaiswal}}, \bibinfo {author} {\bibfnamefont {I.}~\bibnamefont {Procaccia}},
  \bibinfo {author} {\bibfnamefont {C.}~\bibnamefont {Rainone}},\ and\ \bibinfo
  {author} {\bibfnamefont {M.}~\bibnamefont {Singh}},\ }\bibfield  {title}
  {\bibinfo {title} {Mechanical yield in amorphous solids: A first-order phase
  transition},\ }\bibfield  {journal} {\bibinfo  {journal} {Physical Review
  Letters}\ }\textbf {\bibinfo {volume} {116}},\ \href
  {https://doi.org/10.1103/physrevlett.116.085501}
  {10.1103/physrevlett.116.085501} (\bibinfo {year} {2016})\BibitemShut
  {NoStop}%
\bibitem [{\citenamefont {Parisi}\ \emph {et~al.}(2017)\citenamefont {Parisi},
  \citenamefont {Procaccia}, \citenamefont {Rainone},\ and\ \citenamefont
  {Singh}}]{Parisi2017}%
  \BibitemOpen
  \bibfield  {author} {\bibinfo {author} {\bibfnamefont {G.}~\bibnamefont
  {Parisi}}, \bibinfo {author} {\bibfnamefont {I.}~\bibnamefont {Procaccia}},
  \bibinfo {author} {\bibfnamefont {C.}~\bibnamefont {Rainone}},\ and\ \bibinfo
  {author} {\bibfnamefont {M.}~\bibnamefont {Singh}},\ }\bibfield  {title}
  {\bibinfo {title} {Shear bands as manifestation of a criticality in yielding
  amorphous solids},\ }\href {https://doi.org/10.1073/pnas.1700075114}
  {\bibfield  {journal} {\bibinfo  {journal} {Proceedings of the National
  Academy of Sciences}\ }\textbf {\bibinfo {volume} {114}},\ \bibinfo {pages}
  {5577} (\bibinfo {year} {2017})}\BibitemShut {NoStop}%
\bibitem [{\citenamefont {Kobelev}\ and\ \citenamefont
  {Schweizer}(2005)}]{Kobelev2005}%
  \BibitemOpen
  \bibfield  {author} {\bibinfo {author} {\bibfnamefont {V.}~\bibnamefont
  {Kobelev}}\ and\ \bibinfo {author} {\bibfnamefont {K.~S.}\ \bibnamefont
  {Schweizer}},\ }\bibfield  {title} {\bibinfo {title} {Strain softening,
  yielding, and shear thinning in glassy colloidal suspensions},\ }\href
  {https://doi.org/10.1103/PhysRevE.71.021401} {\bibfield  {journal} {\bibinfo
  {journal} {Phys. Rev. E}\ }\textbf {\bibinfo {volume} {71}},\ \bibinfo
  {pages} {021401} (\bibinfo {year} {2005})}\BibitemShut {NoStop}%
\bibitem [{\citenamefont {Nath}\ \emph {et~al.}(2018)\citenamefont {Nath},
  \citenamefont {Ganguly}, \citenamefont {Horbach}, \citenamefont {Sollich},
  \citenamefont {Karmakar},\ and\ \citenamefont {Sengupta}}]{SuSePNAS}%
  \BibitemOpen
  \bibfield  {author} {\bibinfo {author} {\bibfnamefont {P.}~\bibnamefont
  {Nath}}, \bibinfo {author} {\bibfnamefont {S.}~\bibnamefont {Ganguly}},
  \bibinfo {author} {\bibfnamefont {J.}~\bibnamefont {Horbach}}, \bibinfo
  {author} {\bibfnamefont {P.}~\bibnamefont {Sollich}}, \bibinfo {author}
  {\bibfnamefont {S.}~\bibnamefont {Karmakar}},\ and\ \bibinfo {author}
  {\bibfnamefont {S.}~\bibnamefont {Sengupta}},\ }\bibfield  {title} {\bibinfo
  {title} {On the existence of thermodynamically stable rigid solids},\ }\href
  {https://doi.org/10.1073/pnas.1800837115} {\bibfield  {journal} {\bibinfo
  {journal} {Proceedings of the National Academy of Sciences}\ }\textbf
  {\bibinfo {volume} {115}},\ \bibinfo {pages} {E4322} (\bibinfo {year}
  {2018})}\BibitemShut {NoStop}%
\bibitem [{\citenamefont {Lauridsen}\ \emph {et~al.}(2002)\citenamefont
  {Lauridsen}, \citenamefont {Twardos},\ and\ \citenamefont
  {Dennin}}]{LaurisdenFoamPRL2002}%
  \BibitemOpen
  \bibfield  {author} {\bibinfo {author} {\bibfnamefont {J.}~\bibnamefont
  {Lauridsen}}, \bibinfo {author} {\bibfnamefont {M.}~\bibnamefont {Twardos}},\
  and\ \bibinfo {author} {\bibfnamefont {M.}~\bibnamefont {Dennin}},\
  }\bibfield  {title} {\bibinfo {title} {Shear-induced stress relaxation in a
  two-dimensional wet foam},\ }\href
  {https://doi.org/10.1103/PhysRevLett.89.098303} {\bibfield  {journal}
  {\bibinfo  {journal} {Phys. Rev. Lett.}\ }\textbf {\bibinfo {volume} {89}},\
  \bibinfo {pages} {098303} (\bibinfo {year} {2002})}\BibitemShut {NoStop}%
\bibitem [{\citenamefont {Greer}\ \emph {et~al.}(2013)\citenamefont {Greer},
  \citenamefont {Cheng},\ and\ \citenamefont {Ma}}]{GREER201371}%
  \BibitemOpen
  \bibfield  {author} {\bibinfo {author} {\bibfnamefont {A.}~\bibnamefont
  {Greer}}, \bibinfo {author} {\bibfnamefont {Y.}~\bibnamefont {Cheng}},\ and\
  \bibinfo {author} {\bibfnamefont {E.}~\bibnamefont {Ma}},\ }\bibfield
  {title} {\bibinfo {title} {Shear bands in metallic glasses},\ }\href
  {https://doi.org/https://doi.org/10.1016/j.mser.2013.04.001} {\bibfield
  {journal} {\bibinfo  {journal} {Materials Science and Engineering: R:
  Reports}\ }\textbf {\bibinfo {volume} {74}},\ \bibinfo {pages} {71} (\bibinfo
  {year} {2013})}\BibitemShut {NoStop}%
\bibitem [{\citenamefont {Rossi}\ \emph {et~al.}(2022)\citenamefont {Rossi},
  \citenamefont {Biroli}, \citenamefont {Ozawa}, \citenamefont {Tarjus},\ and\
  \citenamefont {Zamponi}}]{RossiTarju2022}%
  \BibitemOpen
  \bibfield  {author} {\bibinfo {author} {\bibfnamefont {S.}~\bibnamefont
  {Rossi}}, \bibinfo {author} {\bibfnamefont {G.}~\bibnamefont {Biroli}},
  \bibinfo {author} {\bibfnamefont {M.}~\bibnamefont {Ozawa}}, \bibinfo
  {author} {\bibfnamefont {G.}~\bibnamefont {Tarjus}},\ and\ \bibinfo {author}
  {\bibfnamefont {F.}~\bibnamefont {Zamponi}},\ }\bibfield  {title} {\bibinfo
  {title} {Finite-disorder critical point in the yielding transition of
  elastoplastic models},\ }\href
  {https://doi.org/10.1103/PhysRevLett.129.228002} {\bibfield  {journal}
  {\bibinfo  {journal} {Phys. Rev. Lett.}\ }\textbf {\bibinfo {volume} {129}},\
  \bibinfo {pages} {228002} (\bibinfo {year} {2022})}\BibitemShut {NoStop}%
\bibitem [{\citenamefont {Singh}\ \emph {et~al.}(2020)\citenamefont {Singh},
  \citenamefont {Ozawa},\ and\ \citenamefont {Berthier}}]{Singh2020}%
  \BibitemOpen
  \bibfield  {author} {\bibinfo {author} {\bibfnamefont {M.}~\bibnamefont
  {Singh}}, \bibinfo {author} {\bibfnamefont {M.}~\bibnamefont {Ozawa}},\ and\
  \bibinfo {author} {\bibfnamefont {L.}~\bibnamefont {Berthier}},\ }\bibfield
  {title} {\bibinfo {title} {Brittle yielding of amorphous solids at finite
  shear rates},\ }\bibfield  {journal} {\bibinfo  {journal} {Physical Review
  Materials}\ }\textbf {\bibinfo {volume} {4}},\ \href
  {https://doi.org/10.1103/physrevmaterials.4.025603}
  {10.1103/physrevmaterials.4.025603} (\bibinfo {year} {2020})\BibitemShut
  {NoStop}%
\bibitem [{\citenamefont {Barlow}\ \emph {et~al.}(2020)\citenamefont {Barlow},
  \citenamefont {Cochran},\ and\ \citenamefont {Fielding}}]{Fielding2020PRL}%
  \BibitemOpen
  \bibfield  {author} {\bibinfo {author} {\bibfnamefont {H.~J.}\ \bibnamefont
  {Barlow}}, \bibinfo {author} {\bibfnamefont {J.~O.}\ \bibnamefont
  {Cochran}},\ and\ \bibinfo {author} {\bibfnamefont {S.~M.}\ \bibnamefont
  {Fielding}},\ }\bibfield  {title} {\bibinfo {title} {Ductile and brittle
  yielding in thermal and athermal amorphous materials},\ }\href
  {https://doi.org/10.1103/PhysRevLett.125.168003} {\bibfield  {journal}
  {\bibinfo  {journal} {Phys. Rev. Lett.}\ }\textbf {\bibinfo {volume} {125}},\
  \bibinfo {pages} {168003} (\bibinfo {year} {2020})}\BibitemShut {NoStop}%
\bibitem [{\citenamefont {Pollard}\ and\ \citenamefont
  {Fielding}(2022)}]{FieldingPRR}%
  \BibitemOpen
  \bibfield  {author} {\bibinfo {author} {\bibfnamefont {J.}~\bibnamefont
  {Pollard}}\ and\ \bibinfo {author} {\bibfnamefont {S.~M.}\ \bibnamefont
  {Fielding}},\ }\bibfield  {title} {\bibinfo {title} {Yielding, shear banding,
  and brittle failure of amorphous materials},\ }\href
  {https://doi.org/10.1103/PhysRevResearch.4.043037} {\bibfield  {journal}
  {\bibinfo  {journal} {Phys. Rev. Res.}\ }\textbf {\bibinfo {volume} {4}},\
  \bibinfo {pages} {043037} (\bibinfo {year} {2022})}\BibitemShut {NoStop}%
\bibitem [{\citenamefont {Richard}\ \emph {et~al.}(2021)\citenamefont
  {Richard}, \citenamefont {Rainone},\ and\ \citenamefont
  {Lerner}}]{LernerBrittle}%
  \BibitemOpen
  \bibfield  {author} {\bibinfo {author} {\bibfnamefont {D.}~\bibnamefont
  {Richard}}, \bibinfo {author} {\bibfnamefont {C.}~\bibnamefont {Rainone}},\
  and\ \bibinfo {author} {\bibfnamefont {E.}~\bibnamefont {Lerner}},\
  }\bibfield  {title} {\bibinfo {title} {{Finite-size study of the athermal
  quasistatic yielding transition in structural glasses}},\ }\bibfield
  {journal} {\bibinfo  {journal} {The Journal of Chemical Physics}\ }\textbf
  {\bibinfo {volume} {155}},\ \href {https://doi.org/10.1063/5.0053303}
  {10.1063/5.0053303} (\bibinfo {year} {2021}),\ \bibinfo {note}
  {056101}\BibitemShut {NoStop}%
\bibitem [{\citenamefont {Greer}(1995)}]{GreerScience}%
  \BibitemOpen
  \bibfield  {author} {\bibinfo {author} {\bibfnamefont {A.~L.}\ \bibnamefont
  {Greer}},\ }\bibfield  {title} {\bibinfo {title} {Metallic glasses},\ }\href
  {https://doi.org/10.1126/science.267.5206.1947} {\bibfield  {journal}
  {\bibinfo  {journal} {Science}\ }\textbf {\bibinfo {volume} {267}},\ \bibinfo
  {pages} {1947} (\bibinfo {year} {1995})}\BibitemShut {NoStop}%
\bibitem [{\citenamefont {Rainone}\ \emph {et~al.}(2015)\citenamefont
  {Rainone}, \citenamefont {Urbani}, \citenamefont {Yoshino},\ and\
  \citenamefont {Zamponi}}]{RainonePRL2015}%
  \BibitemOpen
  \bibfield  {author} {\bibinfo {author} {\bibfnamefont {C.}~\bibnamefont
  {Rainone}}, \bibinfo {author} {\bibfnamefont {P.}~\bibnamefont {Urbani}},
  \bibinfo {author} {\bibfnamefont {H.}~\bibnamefont {Yoshino}},\ and\ \bibinfo
  {author} {\bibfnamefont {F.}~\bibnamefont {Zamponi}},\ }\bibfield  {title}
  {\bibinfo {title} {Following the evolution of hard sphere glasses in infinite
  dimensions under external perturbations: Compression and shear strain},\
  }\href {https://doi.org/10.1103/PhysRevLett.114.015701} {\bibfield  {journal}
  {\bibinfo  {journal} {Phys. Rev. Lett.}\ }\textbf {\bibinfo {volume} {114}},\
  \bibinfo {pages} {015701} (\bibinfo {year} {2015})}\BibitemShut {NoStop}%
\bibitem [{\citenamefont {Urbani}\ and\ \citenamefont
  {Zamponi}(2017)}]{UrbaniPRL2017}%
  \BibitemOpen
  \bibfield  {author} {\bibinfo {author} {\bibfnamefont {P.}~\bibnamefont
  {Urbani}}\ and\ \bibinfo {author} {\bibfnamefont {F.}~\bibnamefont
  {Zamponi}},\ }\bibfield  {title} {\bibinfo {title} {Shear yielding and shear
  jamming of dense hard sphere glasses},\ }\href
  {https://doi.org/10.1103/PhysRevLett.118.038001} {\bibfield  {journal}
  {\bibinfo  {journal} {Phys. Rev. Lett.}\ }\textbf {\bibinfo {volume} {118}},\
  \bibinfo {pages} {038001} (\bibinfo {year} {2017})}\BibitemShut {NoStop}%
\bibitem [{\citenamefont {Rossi}\ \emph {et~al.}(2023)\citenamefont {Rossi},
  \citenamefont {Biroli}, \citenamefont {Ozawa},\ and\ \citenamefont
  {Tarjus}}]{rossi2023far-from-equilibrium-5f0}%
  \BibitemOpen
  \bibfield  {author} {\bibinfo {author} {\bibfnamefont {S.}~\bibnamefont
  {Rossi}}, \bibinfo {author} {\bibfnamefont {G.}~\bibnamefont {Biroli}},
  \bibinfo {author} {\bibfnamefont {M.}~\bibnamefont {Ozawa}},\ and\ \bibinfo
  {author} {\bibfnamefont {G.}~\bibnamefont {Tarjus}},\ }\bibfield  {title}
  {\bibinfo {title} {Far-from-equilibrium criticality in the random-field ising
  model with eshelby interactions},\ }\href
  {https://doi.org/10.1103/physrevb.108.l220202} {\bibfield  {journal}
  {\bibinfo  {journal} {Physical Review B}\ }\textbf {\bibinfo {volume}
  {108}},\ \bibinfo {pages} {L220202} (\bibinfo {year} {2023})},\ \Eprint
  {https://arxiv.org/abs/2309.14919} {2309.14919} \BibitemShut {NoStop}%
\bibitem [{\citenamefont {Mutneja}\ \emph {et~al.}(2025)\citenamefont
  {Mutneja}, \citenamefont {Bhowmik},\ and\ \citenamefont
  {Karmakar}}]{mutnejaBPB2025}%
  \BibitemOpen
  \bibfield  {author} {\bibinfo {author} {\bibfnamefont {A.}~\bibnamefont
  {Mutneja}}, \bibinfo {author} {\bibfnamefont {B.~P.}\ \bibnamefont
  {Bhowmik}},\ and\ \bibinfo {author} {\bibfnamefont {S.}~\bibnamefont
  {Karmakar}},\ }\href {https://arxiv.org/abs/2501.08511} {\bibinfo {title}
  {Finite disorder critical point in the brittle-to-ductile transition of
  amorphous solids in the presence of particle pinning}} (\bibinfo {year}
  {2025}),\ \Eprint {https://arxiv.org/abs/2501.08511} {arXiv:2501.08511
  [cond-mat.soft]} \BibitemShut {NoStop}%
\bibitem [{\citenamefont {Gokhale}\ \emph {et~al.}(2014)\citenamefont
  {Gokhale}, \citenamefont {Hima~Nagamanasa}, \citenamefont {Ganapathy},\ and\
  \citenamefont {Sood}}]{gokhale2014growing}%
  \BibitemOpen
  \bibfield  {author} {\bibinfo {author} {\bibfnamefont {S.}~\bibnamefont
  {Gokhale}}, \bibinfo {author} {\bibfnamefont {K.}~\bibnamefont
  {Hima~Nagamanasa}}, \bibinfo {author} {\bibfnamefont {R.}~\bibnamefont
  {Ganapathy}},\ and\ \bibinfo {author} {\bibfnamefont {A.~K.}\ \bibnamefont
  {Sood}},\ }\bibfield  {title} {\bibinfo {title} {Growing dynamical
  facilitation on approaching the random pinning colloidal glass transition},\
  }\href {https://doi.org/10.1038/ncomms5685} {\bibfield  {journal} {\bibinfo
  {journal} {Nature Communications}\ }\textbf {\bibinfo {volume} {5}},\
  \bibinfo {pages} {4685} (\bibinfo {year} {2014})}\BibitemShut {NoStop}%
\bibitem [{\citenamefont {Das}\ \emph {et~al.}(2023)\citenamefont {Das},
  \citenamefont {Bhowmik}, \citenamefont {Puthirath}, \citenamefont
  {Narayanan},\ and\ \citenamefont {Karmakar}}]{das2023soft}%
  \BibitemOpen
  \bibfield  {author} {\bibinfo {author} {\bibfnamefont {R.}~\bibnamefont
  {Das}}, \bibinfo {author} {\bibfnamefont {B.~P.}\ \bibnamefont {Bhowmik}},
  \bibinfo {author} {\bibfnamefont {A.~B.}\ \bibnamefont {Puthirath}}, \bibinfo
  {author} {\bibfnamefont {T.~N.}\ \bibnamefont {Narayanan}},\ and\ \bibinfo
  {author} {\bibfnamefont {S.}~\bibnamefont {Karmakar}},\ }\bibfield  {title}
  {\bibinfo {title} {Soft pinning: Experimental validation of static
  correlations in supercooled molecular glass-forming liquids},\ }\href@noop {}
  {\bibfield  {journal} {\bibinfo  {journal} {PNAS nexus}\ }\textbf {\bibinfo
  {volume} {2}},\ \bibinfo {pages} {pgad277} (\bibinfo {year}
  {2023})}\BibitemShut {NoStop}%
\bibitem [{\citenamefont {Anwar}\ \emph {et~al.}(2024)\citenamefont {Anwar},
  \citenamefont {Patel}, \citenamefont {Sharma},\ and\ \citenamefont
  {Maitra~Bhattacharyya}}]{anwar2024exploring}%
  \BibitemOpen
  \bibfield  {author} {\bibinfo {author} {\bibfnamefont {E.}~\bibnamefont
  {Anwar}}, \bibinfo {author} {\bibfnamefont {P.}~\bibnamefont {Patel}},
  \bibinfo {author} {\bibfnamefont {M.}~\bibnamefont {Sharma}},\ and\ \bibinfo
  {author} {\bibfnamefont {S.}~\bibnamefont {Maitra~Bhattacharyya}},\
  }\bibfield  {title} {\bibinfo {title} {Exploring the soft pinning effect in
  the dynamics and the structure--dynamics correlation in multicomponent
  supercooled liquids},\ }\href@noop {} {\bibfield  {journal} {\bibinfo
  {journal} {The Journal of Chemical Physics}\ }\textbf {\bibinfo {volume}
  {161}} (\bibinfo {year} {2024})}\BibitemShut {NoStop}%
\bibitem [{\citenamefont {Meenakshi}\ and\ \citenamefont
  {Gupta}(2024)}]{meenakshi2024effect}%
  \BibitemOpen
  \bibfield  {author} {\bibinfo {author} {\bibfnamefont {L.}~\bibnamefont
  {Meenakshi}}\ and\ \bibinfo {author} {\bibfnamefont {B.~S.}\ \bibnamefont
  {Gupta}},\ }\bibfield  {title} {\bibinfo {title} {Effect of annealed disorder
  on the plasticity of amorphous solids},\ }\href@noop {} {\bibfield  {journal}
  {\bibinfo  {journal} {arXiv e-prints}\ ,\ \bibinfo {pages} {arXiv}} (\bibinfo
  {year} {2024})}\BibitemShut {NoStop}%
\bibitem [{\citenamefont {J.}\ \emph {et~al.}(2013)\citenamefont {J.},
  \citenamefont {Bouchbinder},\ and\ \citenamefont
  {Procaccia}}]{ItamarEranJoyPRE2013}%
  \BibitemOpen
  \bibfield  {author} {\bibinfo {author} {\bibfnamefont {A.}~\bibnamefont
  {J.}}, \bibinfo {author} {\bibfnamefont {E.}~\bibnamefont {Bouchbinder}},\
  and\ \bibinfo {author} {\bibfnamefont {I.}~\bibnamefont {Procaccia}},\
  }\bibfield  {title} {\bibinfo {title} {Cooling-rate dependence of the shear
  modulus of amorphous solids},\ }\href
  {https://doi.org/10.1103/PhysRevE.87.042310} {\bibfield  {journal} {\bibinfo
  {journal} {Phys. Rev. E}\ }\textbf {\bibinfo {volume} {87}},\ \bibinfo
  {pages} {042310} (\bibinfo {year} {2013})}\BibitemShut {NoStop}%
\bibitem [{\citenamefont {Grigera}\ and\ \citenamefont
  {Parisi}(2001)}]{MCGrigeraParisi}%
  \BibitemOpen
  \bibfield  {author} {\bibinfo {author} {\bibfnamefont {T.~S.}\ \bibnamefont
  {Grigera}}\ and\ \bibinfo {author} {\bibfnamefont {G.}~\bibnamefont
  {Parisi}},\ }\bibfield  {title} {\bibinfo {title} {Fast monte carlo algorithm
  for supercooled soft spheres},\ }\href
  {https://doi.org/10.1103/PhysRevE.63.045102} {\bibfield  {journal} {\bibinfo
  {journal} {Phys. Rev. E}\ }\textbf {\bibinfo {volume} {63}},\ \bibinfo
  {pages} {045102} (\bibinfo {year} {2001})}\BibitemShut {NoStop}%
\bibitem [{\citenamefont {Gutiérrez}\ \emph {et~al.}(2015)\citenamefont
  {Gutiérrez}, \citenamefont {Karmakar}, \citenamefont {Pollack},\ and\
  \citenamefont {Procaccia}}]{SmarajitSwapTernary}%
  \BibitemOpen
  \bibfield  {author} {\bibinfo {author} {\bibfnamefont {R.}~\bibnamefont
  {Gutiérrez}}, \bibinfo {author} {\bibfnamefont {S.}~\bibnamefont
  {Karmakar}}, \bibinfo {author} {\bibfnamefont {Y.~G.}\ \bibnamefont
  {Pollack}},\ and\ \bibinfo {author} {\bibfnamefont {I.}~\bibnamefont
  {Procaccia}},\ }\bibfield  {title} {\bibinfo {title} {The static lengthscale
  characterizing the glass transition at lower temperatures},\ }\href
  {https://doi.org/10.1209/0295-5075/111/56009} {\bibfield  {journal} {\bibinfo
   {journal} {Europhysics Letters}\ }\textbf {\bibinfo {volume} {111}},\
  \bibinfo {pages} {56009} (\bibinfo {year} {2015})}\BibitemShut {NoStop}%
\bibitem [{\citenamefont {Bhowmik}\ \emph {et~al.}(2020)\citenamefont
  {Bhowmik}, \citenamefont {Ilyin},\ and\ \citenamefont
  {Procaccia}}]{Bhowmik2020}%
  \BibitemOpen
  \bibfield  {author} {\bibinfo {author} {\bibfnamefont {B.~P.}\ \bibnamefont
  {Bhowmik}}, \bibinfo {author} {\bibfnamefont {V.}~\bibnamefont {Ilyin}},\
  and\ \bibinfo {author} {\bibfnamefont {I.}~\bibnamefont {Procaccia}},\
  }\bibfield  {title} {\bibinfo {title} {Thermodynamic equivalence of cyclic
  shear and deep cooling in glass formers},\ }\href
  {https://doi.org/10.1103/PhysRevE.102.010603} {\bibfield  {journal} {\bibinfo
   {journal} {Phys. Rev. E}\ }\textbf {\bibinfo {volume} {102}},\ \bibinfo
  {pages} {010603} (\bibinfo {year} {2020})}\BibitemShut {NoStop}%
\bibitem [{\citenamefont {Bhaumik}\ \emph {et~al.}(2021)\citenamefont
  {Bhaumik}, \citenamefont {Foffi},\ and\ \citenamefont
  {Sastry}}]{Bhaumik2021}%
  \BibitemOpen
  \bibfield  {author} {\bibinfo {author} {\bibfnamefont {H.}~\bibnamefont
  {Bhaumik}}, \bibinfo {author} {\bibfnamefont {G.}~\bibnamefont {Foffi}},\
  and\ \bibinfo {author} {\bibfnamefont {S.}~\bibnamefont {Sastry}},\
  }\bibfield  {title} {\bibinfo {title} {The role of annealing in determining
  the yielding behavior of glasses under cyclic shear deformation},\ }\bibfield
   {journal} {\bibinfo  {journal} {Proceedings of the National Academy of
  Sciences}\ }\textbf {\bibinfo {volume} {118}},\ \href
  {https://doi.org/10.1073/pnas.2100227118} {10.1073/pnas.2100227118} (\bibinfo
  {year} {2021})\BibitemShut {NoStop}%
\bibitem [{\citenamefont {Krishnan}\ \emph {et~al.}(2023)\citenamefont
  {Krishnan}, \citenamefont {Ramola},\ and\ \citenamefont
  {Karmakar}}]{PhysRevApplied.19.024004}%
  \BibitemOpen
  \bibfield  {author} {\bibinfo {author} {\bibfnamefont {V.~V.}\ \bibnamefont
  {Krishnan}}, \bibinfo {author} {\bibfnamefont {K.}~\bibnamefont {Ramola}},\
  and\ \bibinfo {author} {\bibfnamefont {S.}~\bibnamefont {Karmakar}},\
  }\bibfield  {title} {\bibinfo {title} {Annealing effects of multidirectional
  oscillatory shear in model glass formers},\ }\href
  {https://doi.org/10.1103/PhysRevApplied.19.024004} {\bibfield  {journal}
  {\bibinfo  {journal} {Phys. Rev. Appl.}\ }\textbf {\bibinfo {volume} {19}},\
  \bibinfo {pages} {024004} (\bibinfo {year} {2023})}\BibitemShut {NoStop}%
\bibitem [{\citenamefont {Yeh}\ \emph {et~al.}(2020)\citenamefont {Yeh},
  \citenamefont {Ozawa}, \citenamefont {Miyazaki}, \citenamefont {Kawasaki},\
  and\ \citenamefont {Berthier}}]{LudovicCyclicShear}%
  \BibitemOpen
  \bibfield  {author} {\bibinfo {author} {\bibfnamefont {W.-T.}\ \bibnamefont
  {Yeh}}, \bibinfo {author} {\bibfnamefont {M.}~\bibnamefont {Ozawa}}, \bibinfo
  {author} {\bibfnamefont {K.}~\bibnamefont {Miyazaki}}, \bibinfo {author}
  {\bibfnamefont {T.}~\bibnamefont {Kawasaki}},\ and\ \bibinfo {author}
  {\bibfnamefont {L.}~\bibnamefont {Berthier}},\ }\bibfield  {title} {\bibinfo
  {title} {Glass stability changes the nature of yielding under oscillatory
  shear},\ }\href {https://doi.org/10.1103/PhysRevLett.124.225502} {\bibfield
  {journal} {\bibinfo  {journal} {Phys. Rev. Lett.}\ }\textbf {\bibinfo
  {volume} {124}},\ \bibinfo {pages} {225502} (\bibinfo {year}
  {2020})}\BibitemShut {NoStop}%
\bibitem [{\citenamefont {Ozawa}\ \emph {et~al.}(2023)\citenamefont {Ozawa},
  \citenamefont {Iwashita}, \citenamefont {Kob},\ and\ \citenamefont
  {Zamponi}}]{Ozawa2023}%
  \BibitemOpen
  \bibfield  {author} {\bibinfo {author} {\bibfnamefont {M.}~\bibnamefont
  {Ozawa}}, \bibinfo {author} {\bibfnamefont {Y.}~\bibnamefont {Iwashita}},
  \bibinfo {author} {\bibfnamefont {W.}~\bibnamefont {Kob}},\ and\ \bibinfo
  {author} {\bibfnamefont {F.}~\bibnamefont {Zamponi}},\ }\bibfield  {title}
  {\bibinfo {title} {Creating bulk ultrastable glasses by random particle
  bonding},\ }\bibfield  {journal} {\bibinfo  {journal} {Nature
  Communications}\ }\textbf {\bibinfo {volume} {14}},\ \href
  {https://doi.org/10.1038/s41467-023-35812-w} {10.1038/s41467-023-35812-w}
  (\bibinfo {year} {2023})\BibitemShut {NoStop}%
\bibitem [{\citenamefont {Ediger}(2017)}]{EdigerVD}%
  \BibitemOpen
  \bibfield  {author} {\bibinfo {author} {\bibfnamefont {M.~D.}\ \bibnamefont
  {Ediger}},\ }\bibfield  {title} {\bibinfo {title} {{Perspective: Highly
  stable vapor-deposited glasses}},\ }\bibfield  {journal} {\bibinfo  {journal}
  {The Journal of Chemical Physics}\ }\textbf {\bibinfo {volume} {147}},\ \href
  {https://doi.org/10.1063/1.5006265} {10.1063/1.5006265} (\bibinfo {year}
  {2017}),\ \bibinfo {note} {210901}\BibitemShut {NoStop}%
\bibitem [{\citenamefont {Garrett}\ \emph {et~al.}(2012)\citenamefont
  {Garrett}, \citenamefont {Demetriou}, \citenamefont {Chen},\ and\
  \citenamefont {Johnson}}]{Garrett2012}%
  \BibitemOpen
  \bibfield  {author} {\bibinfo {author} {\bibfnamefont {G.~R.}\ \bibnamefont
  {Garrett}}, \bibinfo {author} {\bibfnamefont {M.~D.}\ \bibnamefont
  {Demetriou}}, \bibinfo {author} {\bibfnamefont {J.}~\bibnamefont {Chen}},\
  and\ \bibinfo {author} {\bibfnamefont {W.~L.}\ \bibnamefont {Johnson}},\
  }\bibfield  {title} {\bibinfo {title} {Effect of microalloying on the
  toughness of metallic glasses},\ }\href {https://doi.org/10.1063/1.4769997}
  {\bibfield  {journal} {\bibinfo  {journal} {Applied Physics Letters}\
  }\textbf {\bibinfo {volume} {101}},\ \bibinfo {pages} {241913} (\bibinfo
  {year} {2012})}\BibitemShut {NoStop}%
\bibitem [{\citenamefont {Gonz{\'{a}}lez}(2015)}]{Gonzlez2015}%
  \BibitemOpen
  \bibfield  {author} {\bibinfo {author} {\bibfnamefont {S.}~\bibnamefont
  {Gonz{\'{a}}lez}},\ }\bibfield  {title} {\bibinfo {title} {Role of minor
  additions on metallic glasses and composites},\ }\href
  {https://doi.org/10.1557/jmr.2015.319} {\bibfield  {journal} {\bibinfo
  {journal} {Journal of Materials Research}\ }\textbf {\bibinfo {volume}
  {31}},\ \bibinfo {pages} {76} (\bibinfo {year} {2015})}\BibitemShut {NoStop}%
\bibitem [{\citenamefont {Harmon}\ \emph {et~al.}(2007)\citenamefont {Harmon},
  \citenamefont {Demetriou}, \citenamefont {Johnson},\ and\ \citenamefont
  {Samwer}}]{PhysRevLett.99.135502}%
  \BibitemOpen
  \bibfield  {author} {\bibinfo {author} {\bibfnamefont {J.~S.}\ \bibnamefont
  {Harmon}}, \bibinfo {author} {\bibfnamefont {M.~D.}\ \bibnamefont
  {Demetriou}}, \bibinfo {author} {\bibfnamefont {W.~L.}\ \bibnamefont
  {Johnson}},\ and\ \bibinfo {author} {\bibfnamefont {K.}~\bibnamefont
  {Samwer}},\ }\bibfield  {title} {\bibinfo {title} {Anelastic to plastic
  transition in metallic glass-forming liquids},\ }\href
  {https://doi.org/10.1103/PhysRevLett.99.135502} {\bibfield  {journal}
  {\bibinfo  {journal} {Phys. Rev. Lett.}\ }\textbf {\bibinfo {volume} {99}},\
  \bibinfo {pages} {135502} (\bibinfo {year} {2007})}\BibitemShut {NoStop}%
\bibitem [{\citenamefont {Bhowmik}\ \emph
  {et~al.}(2019{\natexlab{a}})\citenamefont {Bhowmik}, \citenamefont
  {Chaudhuri},\ and\ \citenamefont {Karmakar}}]{Bhowmik2019}%
  \BibitemOpen
  \bibfield  {author} {\bibinfo {author} {\bibfnamefont {B.~P.}\ \bibnamefont
  {Bhowmik}}, \bibinfo {author} {\bibfnamefont {P.}~\bibnamefont {Chaudhuri}},\
  and\ \bibinfo {author} {\bibfnamefont {S.}~\bibnamefont {Karmakar}},\
  }\bibfield  {title} {\bibinfo {title} {Effect of pinning on the yielding
  transition of amorphous solids},\ }\bibfield  {journal} {\bibinfo  {journal}
  {Physical Review Letters}\ }\textbf {\bibinfo {volume} {123}},\ \href
  {https://doi.org/10.1103/physrevlett.123.185501}
  {10.1103/physrevlett.123.185501} (\bibinfo {year}
  {2019}{\natexlab{a}})\BibitemShut {NoStop}%
\bibitem [{\citenamefont {Bhowmik}\ \emph
  {et~al.}(2019{\natexlab{b}})\citenamefont {Bhowmik}, \citenamefont
  {Karmakar}, \citenamefont {Procaccia},\ and\ \citenamefont
  {Rainone}}]{Corrado2019}%
  \BibitemOpen
  \bibfield  {author} {\bibinfo {author} {\bibfnamefont {B.~P.}\ \bibnamefont
  {Bhowmik}}, \bibinfo {author} {\bibfnamefont {S.}~\bibnamefont {Karmakar}},
  \bibinfo {author} {\bibfnamefont {I.}~\bibnamefont {Procaccia}},\ and\
  \bibinfo {author} {\bibfnamefont {C.}~\bibnamefont {Rainone}},\ }\bibfield
  {title} {\bibinfo {title} {Particle pinning suppresses spinodal criticality
  in the shear-banding instability},\ }\href
  {https://doi.org/10.1103/PhysRevE.100.052110} {\bibfield  {journal} {\bibinfo
   {journal} {Phys. Rev. E}\ }\textbf {\bibinfo {volume} {100}},\ \bibinfo
  {pages} {052110} (\bibinfo {year} {2019}{\natexlab{b}})}\BibitemShut
  {NoStop}%
\bibitem [{\citenamefont {Tong}\ and\ \citenamefont {Tanaka}(2019)}]{Tong2019}%
  \BibitemOpen
  \bibfield  {author} {\bibinfo {author} {\bibfnamefont {H.}~\bibnamefont
  {Tong}}\ and\ \bibinfo {author} {\bibfnamefont {H.}~\bibnamefont {Tanaka}},\
  }\bibfield  {title} {\bibinfo {title} {Structural order as a genuine control
  parameter of dynamics in simple glass formers},\ }\bibfield  {journal}
  {\bibinfo  {journal} {Nature Communications}\ }\textbf {\bibinfo {volume}
  {10}},\ \href {https://doi.org/10.1038/s41467-019-13606-3}
  {10.1038/s41467-019-13606-3} (\bibinfo {year} {2019})\BibitemShut {NoStop}%
\bibitem [{\citenamefont {Hima~Nagamanasa}\ \emph {et~al.}(2015)\citenamefont
  {Hima~Nagamanasa}, \citenamefont {Gokhale}, \citenamefont {Sood},\ and\
  \citenamefont {Ganapathy}}]{NagaManasaNatPhys}%
  \BibitemOpen
  \bibfield  {author} {\bibinfo {author} {\bibfnamefont {K.}~\bibnamefont
  {Hima~Nagamanasa}}, \bibinfo {author} {\bibfnamefont {S.}~\bibnamefont
  {Gokhale}}, \bibinfo {author} {\bibfnamefont {A.~K.}\ \bibnamefont {Sood}},\
  and\ \bibinfo {author} {\bibfnamefont {R.}~\bibnamefont {Ganapathy}},\
  }\bibfield  {title} {\bibinfo {title} {Direct measurements of growing
  amorphous order and non-monotonic dynamic correlations in a colloidal
  glass-former},\ }\href {https://doi.org/10.1038/nphys3289} {\bibfield
  {journal} {\bibinfo  {journal} {Nature Physics}\ }\textbf {\bibinfo {volume}
  {11}},\ \bibinfo {pages} {403} (\bibinfo {year} {2015})}\BibitemShut
  {NoStop}%
\bibitem [{\citenamefont {Chikkadi}\ \emph {et~al.}(2011)\citenamefont
  {Chikkadi}, \citenamefont {Wegdam}, \citenamefont {Bonn}, \citenamefont
  {Nienhuis},\ and\ \citenamefont {Schall}}]{ChikkadiPRL2011}%
  \BibitemOpen
  \bibfield  {author} {\bibinfo {author} {\bibfnamefont {V.}~\bibnamefont
  {Chikkadi}}, \bibinfo {author} {\bibfnamefont {G.}~\bibnamefont {Wegdam}},
  \bibinfo {author} {\bibfnamefont {D.}~\bibnamefont {Bonn}}, \bibinfo {author}
  {\bibfnamefont {B.}~\bibnamefont {Nienhuis}},\ and\ \bibinfo {author}
  {\bibfnamefont {P.}~\bibnamefont {Schall}},\ }\bibfield  {title} {\bibinfo
  {title} {Long-range strain correlations in sheared colloidal glasses},\
  }\href {https://doi.org/10.1103/PhysRevLett.107.198303} {\bibfield  {journal}
  {\bibinfo  {journal} {Phys. Rev. Lett.}\ }\textbf {\bibinfo {volume} {107}},\
  \bibinfo {pages} {198303} (\bibinfo {year} {2011})}\BibitemShut {NoStop}%
\bibitem [{\citenamefont {Chikkadi}\ \emph {et~al.}(2014)\citenamefont
  {Chikkadi}, \citenamefont {Miedema}, \citenamefont {Dang}, \citenamefont
  {Nienhuis},\ and\ \citenamefont {Schall}}]{ChikkadiPRL2014}%
  \BibitemOpen
  \bibfield  {author} {\bibinfo {author} {\bibfnamefont {V.}~\bibnamefont
  {Chikkadi}}, \bibinfo {author} {\bibfnamefont {D.~M.}\ \bibnamefont
  {Miedema}}, \bibinfo {author} {\bibfnamefont {M.~T.}\ \bibnamefont {Dang}},
  \bibinfo {author} {\bibfnamefont {B.}~\bibnamefont {Nienhuis}},\ and\
  \bibinfo {author} {\bibfnamefont {P.}~\bibnamefont {Schall}},\ }\bibfield
  {title} {\bibinfo {title} {Shear banding of colloidal glasses: Observation of
  a dynamic first-order transition},\ }\href
  {https://doi.org/10.1103/PhysRevLett.113.208301} {\bibfield  {journal}
  {\bibinfo  {journal} {Phys. Rev. Lett.}\ }\textbf {\bibinfo {volume} {113}},\
  \bibinfo {pages} {208301} (\bibinfo {year} {2014})}\BibitemShut {NoStop}%
\bibitem [{\citenamefont {Mishra}\ \emph {et~al.}(2014)\citenamefont {Mishra},
  \citenamefont {Nagamanasa}, \citenamefont {Ganapathy}, \citenamefont {Sood},\
  and\ \citenamefont {Gokhale}}]{mishraPNASellipsoids2014}%
  \BibitemOpen
  \bibfield  {author} {\bibinfo {author} {\bibfnamefont {C.~K.}\ \bibnamefont
  {Mishra}}, \bibinfo {author} {\bibfnamefont {K.~H.}\ \bibnamefont
  {Nagamanasa}}, \bibinfo {author} {\bibfnamefont {R.}~\bibnamefont
  {Ganapathy}}, \bibinfo {author} {\bibfnamefont {A.~K.}\ \bibnamefont
  {Sood}},\ and\ \bibinfo {author} {\bibfnamefont {S.}~\bibnamefont
  {Gokhale}},\ }\bibfield  {title} {\bibinfo {title} {Dynamical facilitation
  governs glassy dynamics in suspensions of colloidal ellipsoids},\ }\href
  {https://doi.org/10.1073/pnas.1413384111} {\bibfield  {journal} {\bibinfo
  {journal} {Proceedings of the National Academy of Sciences}\ }\textbf
  {\bibinfo {volume} {111}},\ \bibinfo {pages} {15362} (\bibinfo {year}
  {2014})},\ \Eprint
  {https://arxiv.org/abs/https://www.pnas.org/doi/pdf/10.1073/pnas.1413384111}
  {https://www.pnas.org/doi/pdf/10.1073/pnas.1413384111} \BibitemShut {NoStop}%
\bibitem [{\citenamefont {Mishra}\ and\ \citenamefont
  {Ganapathy}(2015)}]{mishraPRL2015ellipsoids}%
  \BibitemOpen
  \bibfield  {author} {\bibinfo {author} {\bibfnamefont {C.~K.}\ \bibnamefont
  {Mishra}}\ and\ \bibinfo {author} {\bibfnamefont {R.}~\bibnamefont
  {Ganapathy}},\ }\bibfield  {title} {\bibinfo {title} {Shape of dynamical
  heterogeneities and fractional stokes-einstein and stokes-einstein-debye
  relations in quasi-two-dimensional suspensions of colloidal ellipsoids},\
  }\href {https://doi.org/10.1103/PhysRevLett.114.198302} {\bibfield  {journal}
  {\bibinfo  {journal} {Phys. Rev. Lett.}\ }\textbf {\bibinfo {volume} {114}},\
  \bibinfo {pages} {198302} (\bibinfo {year} {2015})}\BibitemShut {NoStop}%
\bibitem [{\citenamefont {Kob}\ and\ \citenamefont {Andersen}(1995)}]{KA}%
  \BibitemOpen
  \bibfield  {author} {\bibinfo {author} {\bibfnamefont {W.}~\bibnamefont
  {Kob}}\ and\ \bibinfo {author} {\bibfnamefont {H.~C.}\ \bibnamefont
  {Andersen}},\ }\bibfield  {title} {\bibinfo {title} {Testing mode-coupling
  theory for a supercooled binary lennard-jones mixture i: The van hove
  correlation function},\ }\href {https://doi.org/10.1103/PhysRevE.51.4626}
  {\bibfield  {journal} {\bibinfo  {journal} {Phys. Rev. E}\ }\textbf {\bibinfo
  {volume} {51}},\ \bibinfo {pages} {4626} (\bibinfo {year}
  {1995})}\BibitemShut {NoStop}%
\bibitem [{\citenamefont {Br\"{u}ning}\ \emph {et~al.}(2008)\citenamefont
  {Br\"{u}ning}, \citenamefont {St-Onge}, \citenamefont {Patterson},\ and\
  \citenamefont {Kob}}]{2dmKA}%
  \BibitemOpen
  \bibfield  {author} {\bibinfo {author} {\bibfnamefont {R.}~\bibnamefont
  {Br\"{u}ning}}, \bibinfo {author} {\bibfnamefont {D.~A.}\ \bibnamefont
  {St-Onge}}, \bibinfo {author} {\bibfnamefont {S.}~\bibnamefont {Patterson}},\
  and\ \bibinfo {author} {\bibfnamefont {W.}~\bibnamefont {Kob}},\ }\bibfield
  {title} {\bibinfo {title} {Glass transitions in one-, two-, three-, and
  four-dimensional binary lennard-jones systems},\ }\href
  {https://doi.org/10.1088/0953-8984/21/3/035117} {\bibfield  {journal}
  {\bibinfo  {journal} {Journal of Physics: Condensed Matter}\ }\textbf
  {\bibinfo {volume} {21}},\ \bibinfo {pages} {035117} (\bibinfo {year}
  {2008})}\BibitemShut {NoStop}%
\bibitem [{\citenamefont {Rycroft}\ \emph {et~al.}(2006)\citenamefont
  {Rycroft}, \citenamefont {Grest}, \citenamefont {Landry},\ and\ \citenamefont
  {Bazant}}]{PhysRevE.74.021306}%
  \BibitemOpen
  \bibfield  {author} {\bibinfo {author} {\bibfnamefont {C.~H.}\ \bibnamefont
  {Rycroft}}, \bibinfo {author} {\bibfnamefont {G.~S.}\ \bibnamefont {Grest}},
  \bibinfo {author} {\bibfnamefont {J.~W.}\ \bibnamefont {Landry}},\ and\
  \bibinfo {author} {\bibfnamefont {M.~Z.}\ \bibnamefont {Bazant}},\ }\bibfield
   {title} {\bibinfo {title} {Analysis of granular flow in a pebble-bed nuclear
  reactor},\ }\href {https://doi.org/10.1103/PhysRevE.74.021306} {\bibfield
  {journal} {\bibinfo  {journal} {Phys. Rev. E}\ }\textbf {\bibinfo {volume}
  {74}},\ \bibinfo {pages} {021306} (\bibinfo {year} {2006})}\BibitemShut
  {NoStop}%
\bibitem [{\citenamefont {Gendelman}\ \emph {et~al.}(2014)\citenamefont
  {Gendelman}, \citenamefont {Joy}, \citenamefont {Mishra}, \citenamefont
  {Procaccia},\ and\ \citenamefont {Samwer}}]{Gendelman2014}%
  \BibitemOpen
  \bibfield  {author} {\bibinfo {author} {\bibfnamefont {O.}~\bibnamefont
  {Gendelman}}, \bibinfo {author} {\bibfnamefont {A.}~\bibnamefont {Joy}},
  \bibinfo {author} {\bibfnamefont {P.}~\bibnamefont {Mishra}}, \bibinfo
  {author} {\bibfnamefont {I.}~\bibnamefont {Procaccia}},\ and\ \bibinfo
  {author} {\bibfnamefont {K.}~\bibnamefont {Samwer}},\ }\bibfield  {title}
  {\bibinfo {title} {On the effect of microalloying on the mechanical
  properties of metallic glasses},\ }\href
  {https://doi.org/10.1016/j.actamat.2013.10.029} {\bibfield  {journal}
  {\bibinfo  {journal} {Acta Materialia}\ }\textbf {\bibinfo {volume} {63}},\
  \bibinfo {pages} {209–215} (\bibinfo {year} {2014})}\BibitemShut {NoStop}%
\bibitem [{\citenamefont {Liu}\ \emph {et~al.}(2023)\citenamefont {Liu},
  \citenamefont {Liu},\ and\ \citenamefont {Peng}}]{LIU2023122052}%
  \BibitemOpen
  \bibfield  {author} {\bibinfo {author} {\bibfnamefont {Y.}~\bibnamefont
  {Liu}}, \bibinfo {author} {\bibfnamefont {H.}~\bibnamefont {Liu}},\ and\
  \bibinfo {author} {\bibfnamefont {H.}~\bibnamefont {Peng}},\ }\bibfield
  {title} {\bibinfo {title} {Pinning effect on the correlations of nonaffine
  displacement in metallic glasses},\ }\href
  {https://doi.org/https://doi.org/10.1016/j.jnoncrysol.2022.122052} {\bibfield
   {journal} {\bibinfo  {journal} {Journal of Non-Crystalline Solids}\ }\textbf
  {\bibinfo {volume} {601}},\ \bibinfo {pages} {122052} (\bibinfo {year}
  {2023})}\BibitemShut {NoStop}%
\bibitem [{\citenamefont {Falk}\ and\ \citenamefont {Langer}(1998)}]{Falk1998}%
  \BibitemOpen
  \bibfield  {author} {\bibinfo {author} {\bibfnamefont {M.~L.}\ \bibnamefont
  {Falk}}\ and\ \bibinfo {author} {\bibfnamefont {J.~S.}\ \bibnamefont
  {Langer}},\ }\bibfield  {title} {\bibinfo {title} {Dynamics of viscoplastic
  deformation in amorphous solids},\ }\href
  {https://doi.org/10.1103/physreve.57.7192} {\bibfield  {journal} {\bibinfo
  {journal} {Physical Review E}\ }\textbf {\bibinfo {volume} {57}},\ \bibinfo
  {pages} {7192} (\bibinfo {year} {1998})}\BibitemShut {NoStop}%
\bibitem [{\citenamefont {Blackburn}\ \emph {et~al.}(1996)\citenamefont
  {Blackburn}, \citenamefont {Wang},\ and\ \citenamefont
  {Ediger}}]{Blackburn1996}%
  \BibitemOpen
  \bibfield  {author} {\bibinfo {author} {\bibfnamefont {F.~R.}\ \bibnamefont
  {Blackburn}}, \bibinfo {author} {\bibfnamefont {C.-Y.}\ \bibnamefont
  {Wang}},\ and\ \bibinfo {author} {\bibfnamefont {M.~D.}\ \bibnamefont
  {Ediger}},\ }\bibfield  {title} {\bibinfo {title} {Translational and
  rotational motion of probes in supercooled 1, 3, 5-tris(naphthyl)benzene},\
  }\href {https://doi.org/10.1021/jp9622041} {\bibfield  {journal} {\bibinfo
  {journal} {The Journal of Physical Chemistry}\ }\textbf {\bibinfo {volume}
  {100}},\ \bibinfo {pages} {18249} (\bibinfo {year} {1996})}\BibitemShut
  {NoStop}%
\bibitem [{\citenamefont {Mutneja}\ and\ \citenamefont
  {Karmakar}(2021)}]{AnoopRot}%
  \BibitemOpen
  \bibfield  {author} {\bibinfo {author} {\bibfnamefont {A.}~\bibnamefont
  {Mutneja}}\ and\ \bibinfo {author} {\bibfnamefont {S.}~\bibnamefont
  {Karmakar}},\ }\bibfield  {title} {\bibinfo {title} {Probing dynamic
  heterogeneity and amorphous order using rotational dynamics of rodlike
  particles in supercooled liquids},\ }\href
  {https://doi.org/10.1103/PhysRevApplied.16.034022} {\bibfield  {journal}
  {\bibinfo  {journal} {Phys. Rev. Applied}\ }\textbf {\bibinfo {volume}
  {16}},\ \bibinfo {pages} {034022} (\bibinfo {year} {2021})}\BibitemShut
  {NoStop}%
\end{thebibliography}%
\pagebreak
\newpage
\clearpage
\title{Supplementary Information : Finite Disorder Critical Point in Brittle-to-Ductile Transition in Amorphous Solids with Aspherical Impurities}
\maketitle
\beginsupplement
\begin{figure*}
	\centering
	\includegraphics[width=0.55\textwidth]{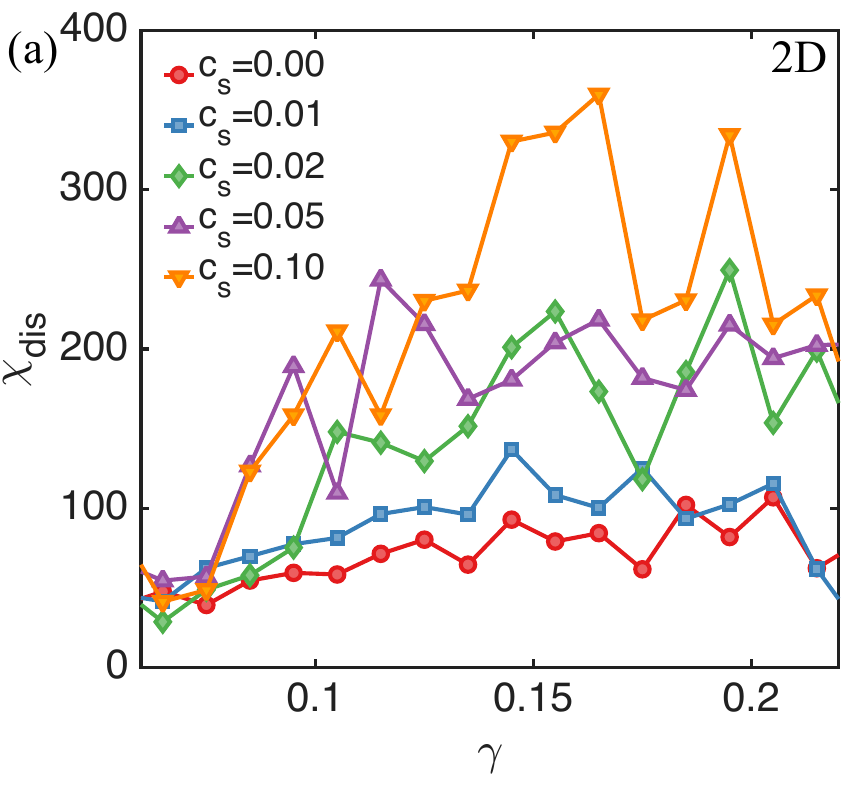}
	\caption{\textbf{Susceptibility plots of the 2dmKA system with varying concentrations of spherical impurities:} This supplemental plot accompanies Fig. 2 from the main text to illustrate the increase in susceptibility (using Eq. (1) from the main text) resulting from the addition of spherical impurities. The increased magnitude of susceptibility indicates an enhanced brittle character.}
\end{figure*}
\begin{figure*}
	\centering
	\includegraphics[width=0.55\textwidth]{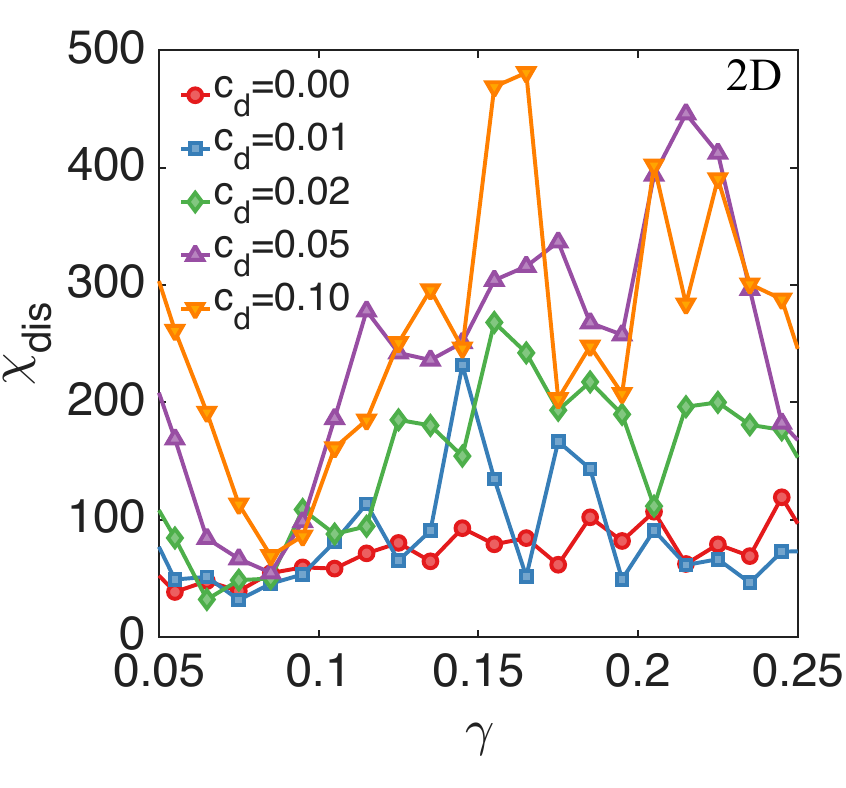}
	\caption{\textbf{Susceptibility plots of the 2dmKA system with different aspherical dimer impurity concentrations:} This is a supplemental plot to Fig. 3 of the main text that shows the increase in susceptibility (using Eq. (1) from the main text) with the addition of aspherical dimer impurities.}
\end{figure*}

\begin{figure*}
	\centering
	\includegraphics[width=0.6\textwidth]{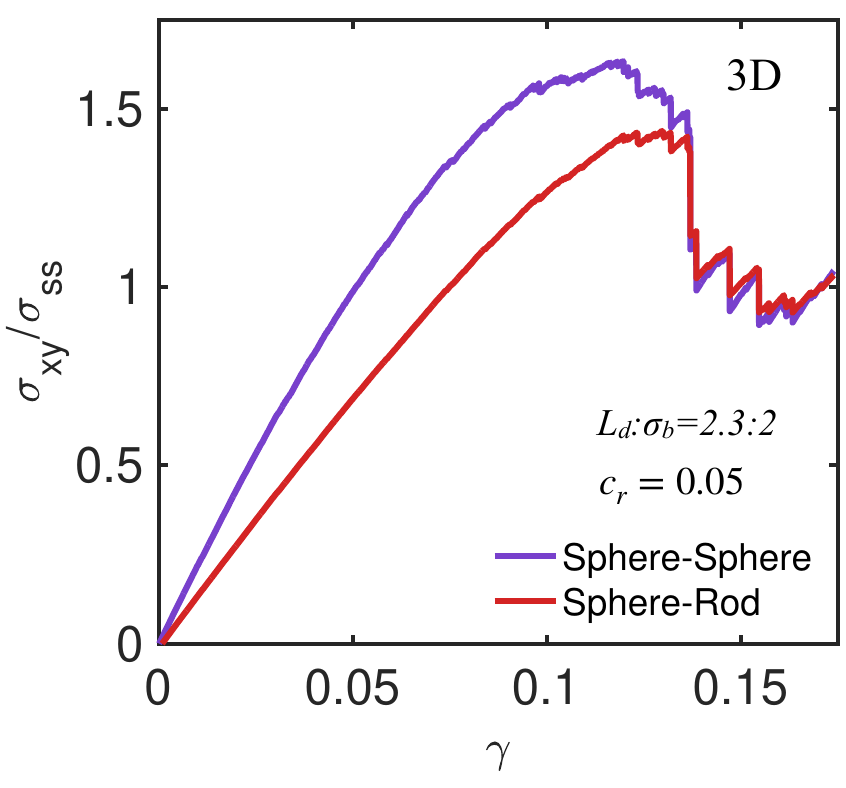}
	\caption{\textbf{Single ensemble stress contribution bisected into sphere-sphere interaction and sphere-rod interaction:} This is a supplemental plot to Fig. 2(d) of the main text to show the one-to-one correspondence between the plastic events of the two stress contributions in a doped system. Moreover, the sphere-sphere interactions start to have much larger plastic drops earlier, while the sphere-rod pairs continue to store stress.}
\end{figure*}
\begin{figure*}
	\centering
	\includegraphics[width=0.7\textwidth]{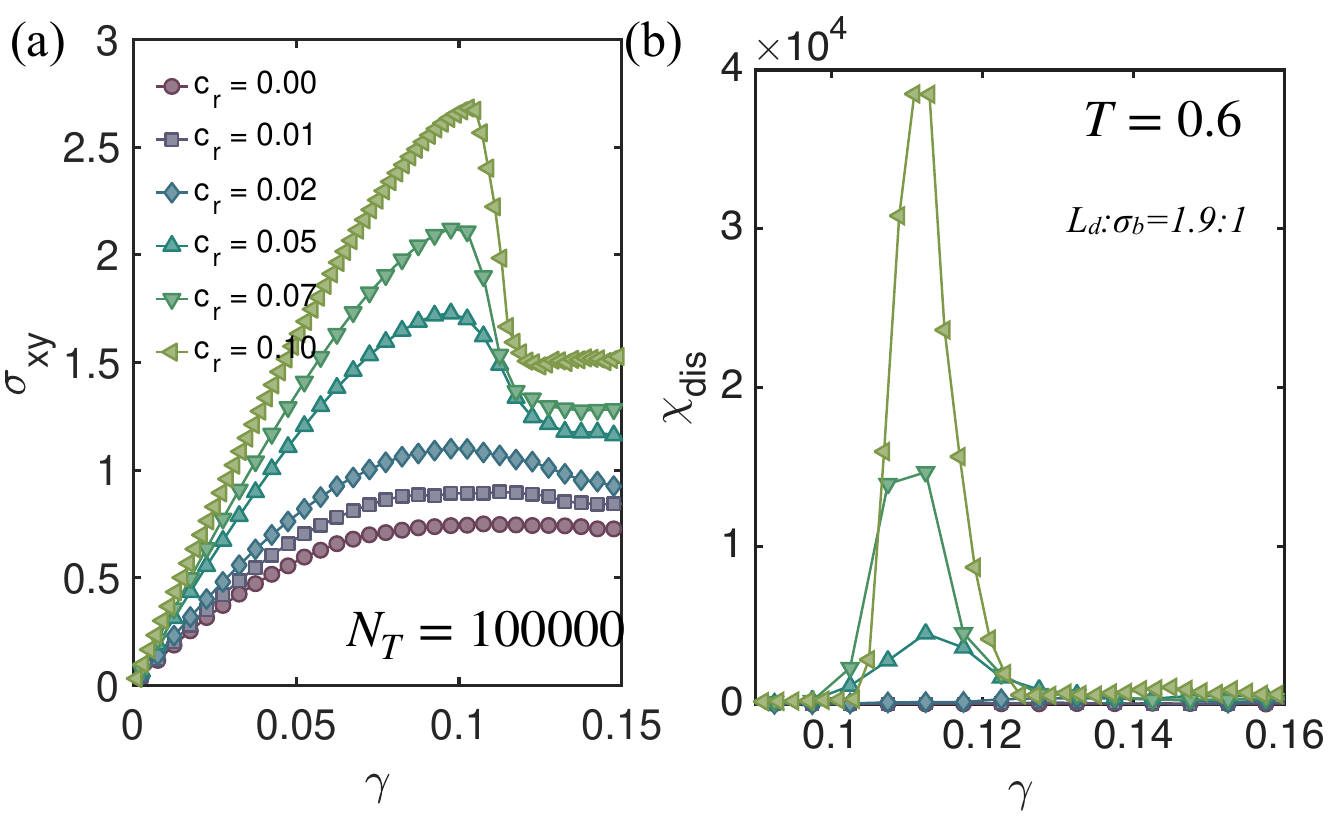}
	\caption{\textbf{Ductile-to-brittle transition with the addition of rotationally frozen rod impurities:} This plot supplements Fig. 7 of the main text by illustrating (a) the averaged stress-strain response shifting from ductile to brittle in high temperature systems with increasing concentrations of rotationally frozen (frozen non-affine rotations) impurities, and (b) their corresponding susceptibility plots.   }
\end{figure*}

\end{document}